\documentclass[preprints,article,accept,moreauthors,pdftex]{mdpi}

\usepackage{siunitx}
\usepackage{physics}
\usepackage{upgreek}
\usepackage{bm}

\firstpage{1} 
\pubvolume{1}
\issuenum{1}
\articlenumber{0}
\pubyear{2021}
\copyrightyear{2020}
\datereceived{} 
\dateaccepted{} 
\datepublished{} 
\hreflink{https://doi.org/} % If needed use \linebreak
\usepackage{amssymb}
\usepackage{physics}
\usepackage{siunitx}
\usepackage{supertabular}
\usepackage{graphicx}
\usepackage{booktabs}
\usepackage{multicol}
\usepackage{hyperref}
\usepackage{caption}
\usepackage{subcaption}
\usepackage{ar}
\usepackage{upgreek}
\usepackage{multirow}
\usepackage{float}
\usepackage{amsmath}

\DeclareSIUnit\rpm{rpm}

\Title{Deep reinforcement learning for flow control exploits different physics for increasing Reynolds-number regimes}

\TitleCitation{Deep reinforcement learning for flow control exploits different physics for increasing Reynolds-number regimes}

\Author{Pau Varela$^{1}$\orcidA{},%
Pol Suárez$^{2,3}$\orcidB{},%
Francisco Alcántara-Ávila$^{3}$\orcidJ{},%
Arnau Miró$^{2}\orcidG{}$,%
Jean Rabault$^{4}$\orcidC{},%
Bernat Font$^{2}$\orcidE{},%
Luis Miguel García-Cuevas$^{1}$\orcidF{},%
Oriol Lehmkuhl$^{2}$\orcidH{}%
and Ricardo Vinuesa$^{3}$*\orcidI{}}

\AuthorNames{Pau Varela Martínez, Pol Suárez, Francisco Alcántara-Ávila, Arnau Miró, Jean Rabault, Bernat Font, Luis Miguel García-Cuevas González, Oriol Lehmkhul, Ricardo Vinuesa
}

\AuthorCitation{Varela, P. et al.}
\address{%
$^{1}$ \quad CMT - Motores Térmicos, Universitat Politècnica de València, Spain

$^{2}$ \quad Barcelona Super Computing Center - Centro Nacional de Supercomputación (BSC-CNS), Spain

$^{3}$ \quad FLOW, Engineering Mechanics, KTH Royal Institute of Technology, Stockholm, Sweden

$^{4}$ \quad Norwegian Meteorological Institute, Norway

}

\corres{Correspondence: rvinuesa@mech.kth.se}

\abstract{The current increase of emissions associated with aviation requires deeper research on novel sensing and flow-control strategies to obtain improved aerodynamic performance. In this context, data-driven methods are suitable for exploring new approaches to control the flow and develop more efficient strategies. Deep artificial neural networks (ANNs) used together with reinforcement learning, \textit{i.e.} deep reinforcement learning (DRL), are receiving growing attention due to their capabilities to control complex problems in multiple areas. In particular, this technique has been recently used to solve problems related to flow control. In this work, an ANN trained through a DRL agent, coupled with the numerical solver Alya, is used to perform active flow control. The Tensorforce library is used to apply DRL to the simulated flow. Two-dimensional simulations of the flow around a cylinder are conducted and an active control based on two jets located on the walls of the cylinder is considered. By gathering information from the flow surrounding the cylinder, the ANN agent is able to learn through a proximal-policy optimisation (PPO) effective control strategies for the jets, leading to a significant drag reduction. Furthermore, the agent needs to account for the coupled effects of the friction- and pressure-drag components, as well as the interaction between the two boundary layers on both sides of the cylinder and the wake. In the present work, a Reynolds-number range beyond those previously considered is studied and compared with results obtained using classical flow-control methods. Significantly different nature in the control strategies is identified by the DRL as the Reynolds number $Re$ increases. On one hand, for $Re \leq 1000$ the classical control strategy based on an opposition control relative to the wake oscillation is obtained. On the other hand, for $Re = 2000$ the new strategy consists of an energisation of the boundary layers and the separation area, which modulate the flow separation and reduce drag in a fashion similar to that of the drag crisis, through a high frequency actuation. A cross-application of agents is performed for a flow at $Re = 2000$, obtaining similar results in terms of drag reduction with the agents trained at $Re = 1000$ and $2000$. The fact that two different strategies yield the same performance make us question whether this Reynolds number regime ($Re = 2000$) belongs to a transition towards a nature-different flow which would only admit a high-frequency actuation strategy to obtain drag reduction. At the same time, this finding allows the application of ANNs trained at lower Reynolds numbers but comparable in nature, saving computational resources.}


\begin{document}
%%%%%%%%%%%%%%%%%%%%%%%%%%%%%%%%%%%%%%%%%%
%\setcounter{section}{-1} %% Remove this when starting to work on the template.

%The order of the section titles is: Introduction, Materials and Methods, Results, Discussion, Conclusions for these journals: aerospace,algorithms,antibodies,antioxidants,atmosphere,axioms,biomedicines,carbon,crystals,designs,diagnostics,environments,fermentation,fluids,forests,fractalfract,informatics,information,inventions,jfmk,jrfm,lubricants,neonatalscreening,neuroglia,particles,pharmaceutics,polymers,processes,technologies,viruses,vision

\section{Introduction}

In transport applications and, especially, in aeronautics, drag reduction is directly related to a decrease in fuel use, which translates into reducing polluting and greenhouse-gas emissions \cite{HOWELL}. Over the past decades, several techniques have been developed and used to reduce drag, both passively (\citet{PASSIVE}) and actively (\citet{ACTIVE}). Passive methods typically rely on fixed geometric changes without using actuators. An example of this technology would be the widespread use of winglets in aircraft. Inspired by bird wings, winglets consist of small wing extensions at the wing tip with an angle relative to the wing-span direction. Using winglets, lift-induced drag is decreased by reducing the size and formation of vortices at the wing tip \cite{GUERRERO}, incurring in (hopefully) small increases in structural weight and parasitic drag. Regarding active methods, diverse techniques are used to reduce aircraft drag and associated emissions. In the work of \citet{TISEIRA} and \citet{SERRANO1,SERRANO2}, the use of distributed electric propulsion is combined with boundary-layer ingestion, setting small propellers along the wing near the trailing edge. Thanks to both technologies, it is possible to increase the aerodynamic efficiency of small aircraft or unmanned air vehicles (UAVs). In the same way, the location and use of both jet pumps in wings and synthetic jets are studied to reduce the drag of different aerodynamic bodies. An example of blowing and suction used to control turbulent boundary layers is provided by \citet{KAMETANI}, and a number of studies have shown the feasibility of this approach in turbulent wings \cite{VINUESA1,VINUESA2,VINUESA3,VINUESA4}.

Additional examples can be found in the the work of \citet{VOEVODIN} and \citet{YOUSEFI}, where the placement of suction and ejection in wing airfoils is studied to reduce aerodynamic drag or achieve efficient control of the flow around the wing in specific flight-operating conditions. In the same way, the use of synthetic jets has been studied to achieve this same objective, as displayed in the work of \citet{CUI} on an Ahmed body, or the investigation of \citet{PARK} using an array of synthetic jets applied to a car. A comprehensive review of active flow control in turbulent flows is provided by \citet{CHOI}.

Active-control methods rely on complicated control strategies due to their variable behaviour and dependencies. As shown by \citet{MUDDADA}, even the control of simple actuators to reduce the drag of a cylinder immersed in a low-Reynolds-number flow can be complicated due to the interaction between boundary layers, separating shear layers and wake. In this case, by correctly developing a control algorithm, the drag is reduced by about \SI{53}{\percent}, leaving ample room for improvement in the case of achieving better control. However, thanks to the application of artificial neural networks (ANNs) and deep reinforcement learning (DRL), it is possible to develop functional control strategies at an affordable computational cost in similar problems, as shown by the work by \citet{RABAULT1}. An ANN is a non-parametric tool formed by layers of connected processing nodes or neurons and it can be trained to solve complex problems \cite{VIQUERAT1, pino2022comparative}. The ANN training can take place through different types of learning, where DRL is one of the fastest-growing in solving a wide range of cases\cite{VIQUERAT2, rabault2020deep},  including flow control in complex geometries \cite{VINUESA5,VINUESA6}.

DRL is based on maximising a reward by means of an agent interacting with the environment through actions, based on partial observations of the environment. Note that the combination of DRL and ANN was successfully carried out to solve active flow control in the work of \citet{RABAULT1}. A proximal-policy-optimisation (PPO) is used to obtain the control policy of two synthetic jets on a two-dimensional cylinder in a low-Reynolds-number flow. PPO parameterises a policy function using an ANN with a set of neuron weights given an observation state. The ANN then produces a set of moments that describe a distribution function from which actions are sampled. It is possible to obtain an expression for the estimation of the gradient of the reward as a function of the neuron weights. Additionally, the PPO uses a critic network that estimates the new reward function, which is helpful with stochastic data. Also, there is a limit in the maximum update at every training step, preventing rare events from producing a large reward \cite{BELUS}. It was shown that DRL was feasible and enabled finding a control strategy such that the cylinder drag was reduced by \SI{8}{\percent}. Rabault and Kuhnle \cite{RABAULT2} developed a framework to parallelise environments, enabling speeding up computation and learning, thus reaching a better solution faster. Additionally, \citet{RABAULT3} validated the DRL approach in a more complex problem using four jets on the cylinder and extending the Reynolds-number range. The results of Rabault \textit{et al.} \cite{RABAULT1,RABAULT2} have been extended by a multitude of studies in recent years. For example, the works of \citet{TOKAREV} or \citet{xu2020active}, where an oscillatory rotary control is applied to the cylinder; the study of linear stability and low sensitivity that allows a better understanding of DRL by \citet{LI}, or the effort of \citet{REN} where the same DRL approach is used with a solver that allows calculations at higher Reynolds numbers, with the additional problem of controlling the increase in turbulence around the cylinder. Applications to square cylinders, which are relevant to civil engineering, have also been presented recently \cite{doi:10.1063/5.0103113}, as well as using traditional modal analysis methods for defining effective reward functions \cite{qin2021application}. Regarding these works, there is room for improvement through the application of high-performance computing (HPCs), enabling conducting faster calculations at higher Reynolds numbers not previously calculated.

The main objective of this work is, indeed, to demonstrate the capabilities of the HPC resources available today to perform aerodynamic optimisation with DRL-based control at much higher Reynolds numbers than the ones already studied in the literature. This project proposes the use of the numerical code Alya, developed in the Barcelona Supercomputer Center (BSC-CNS) \cite{VAZQUEZ}, employing the HPC Marenostrum IV cluster. Alya is a tier-0 massively parallel code and is designed to solve the discretised partial differential equations using finite elements. This code has been successfully used in different problems of fluid mechanics, including simulations of turbulent flows \cite{OWEN,LEHMKUHL}.

In the present work, the methodology is first detailed in \autoref{sec:Methodology}, dedicated to the numerical setup and the application of the DRL and the ANN. The results are then summarised and discussed in \autoref{sec:Results}, which starts with the validation of the Alya solver in the combination of an HPC cluster to solve the DRL problem in the cylinder. Then, the results derived from the increase in Reynolds number, which exceed the conditions reported in the literature, are shown. Finally, the main conclusions are collected in \autoref{sec:Conclusions}.

\section{Methods}\label{sec:Methodology}

This section is divided into two parts: the problem description and the domain setup along the methodology regarding the numerical simulations; and the framework including the DRL algorithm.

\subsection{Problem configuration and numerical setup}\label{sec:prob_setup}

\begin{figure}[H]
\centering
\includegraphics[width=0.75\textwidth]{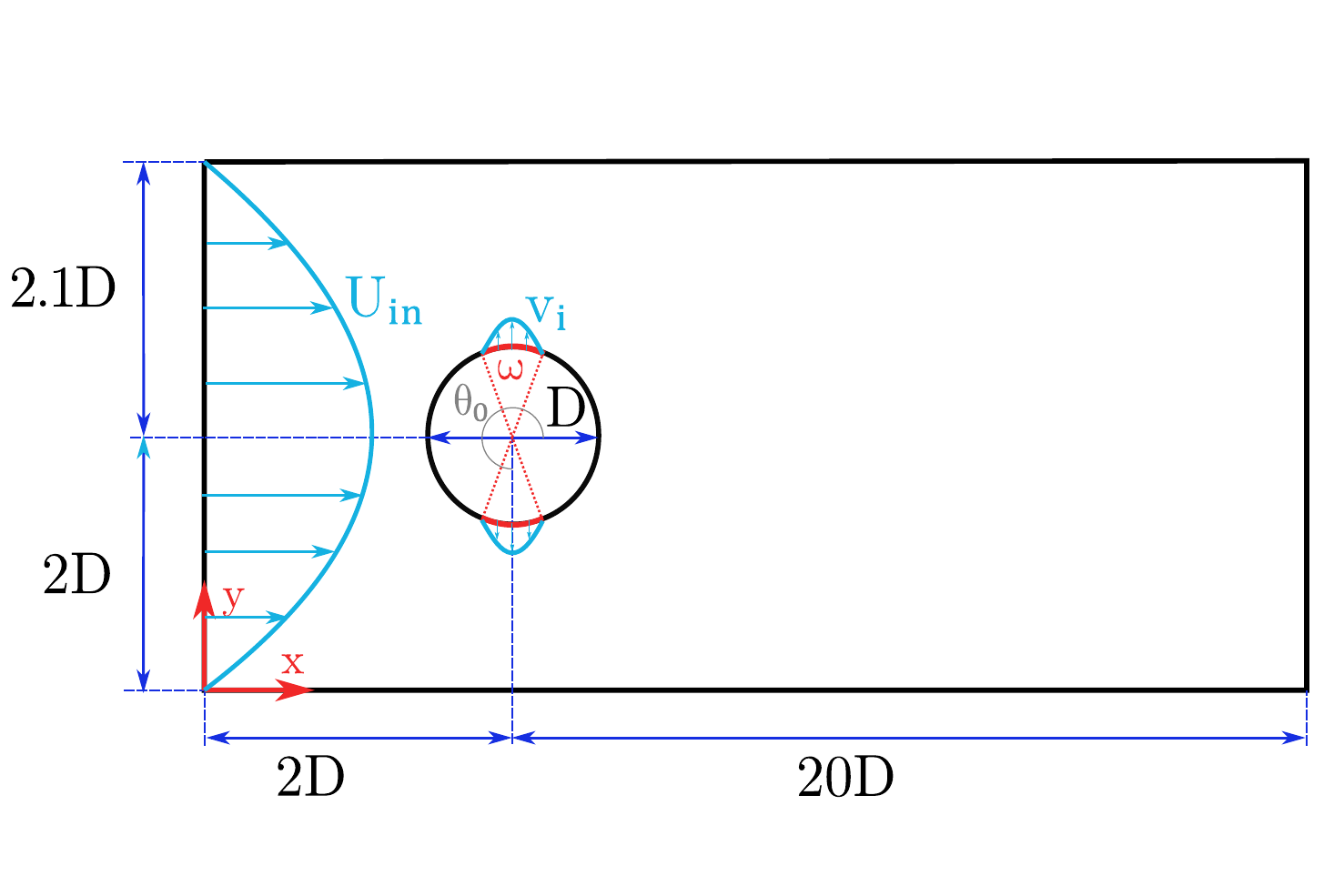}
\caption{Main domain dimensions in terms of the cylinder diameter $D$, where $\omega$ represents the jet width and $\theta_{0}$ is the angle of the center of the jet. In light blue the parabolic velocity profile of the inlet and the jet are represented. The domain representation is not to scale. }
\label{fig:dominio}
\end{figure}

The domain simulation is two dimensional (2D) and consists of a cylinder immersed in a rectangular channel, in the same configuration as that described in the work by \citet{RABAULT1}. The domain is normalised using a cylinder with diameter $D$ as the reference length scale. The channel has a dimension of $L=22$ in length, aligned with the main flow direction, and a height equal to $H=4.1$. The origin of the coordinate system is set on the left-bottom corner of the channel, and the center of the cylinder displaced towards the bottom by $0.05 D$ units, to develop the vortex-shedding behind it. A schematic representation of the domain is depicted in \autoref{fig:dominio}. Aligned with the vertical axis and symmetrically on the cylinder wall, one at the top and one at the bottom, there are two synthetic jets with a total opening of $\omega=\SI{10}{\degree}$ each. This position of the jets is chosen so their actuation is normal to the flow and drag reduction comes from effective actuation rather than injection of momentum.

Regarding the boundary conditions of the problem, an inlet parabolic velocity condition is imposed and expressed as in \autoref{eqn:U_inlet}:

\begin{equation} 
U_\textnormal{in}\pqty{y}=U_\textnormal{max}\bqty{1-\pqty{2\frac{y-H/2}{H}}^2},
  \label{eqn:U_inlet}
\end{equation}

\noindent where $[U_\textnormal{in}(y),V_\textnormal{in}(y)=0]$ is the velocity vector and $U_\textnormal{max}$ is the maximum velocity reached at the middle of the channel, which is equal to \num{1.5} times the mean velocity $\overline{U}$, defined as shown in \autoref{eqn:U_mean}:

\begin{equation} 
\overline{U}=\frac{1}{H}\int^{H}_{0}U_\textnormal{in}(y){\rm d}y=\frac{2}{3}U_\textnormal{max}.
  \label{eqn:U_mean}
\end{equation}

A value of $U_{\rm{max}} = 1.5$ is used so that the scaling velocity of the problem is $\overline{U} = 1$. A no-slip condition is imposed on the cylinder solid wall, while a smooth-wall condition is imposed on the channel walls. The right boundary of the channel is set as a free outlet with zero velocity gradient and constant pressure. The jet velocity ($v_i$) is a function of both the set jet angle ($\theta$) and the mass flow rate ($Q$) determined by the ANN, as described in \autoref{eqn:jet_A}:

\begin{equation} 
v_{i}=Q \frac{\uppi}{2 \omega R^2} \cos\bqty{\frac{\uppi}{\omega}(\theta-\theta_{0})},
  \label{eqn:jet_A}
\end{equation}

\noindent where $\theta_{0}$ corresponds to the angle where the jet is centred and $R$ is the cylinder radius. The scaling factor $\uppi/(2 \omega R^2)$ is used so that the integration of the jet velocity over the jet width gives the desired mass flow rate $Q$. More details about the intensity parameters are provided in \autoref{sec:DRL_setup}.

\begin{figure}[H]
\begin{subfigure}[b]{\linewidth}
\includegraphics[width=1\linewidth]{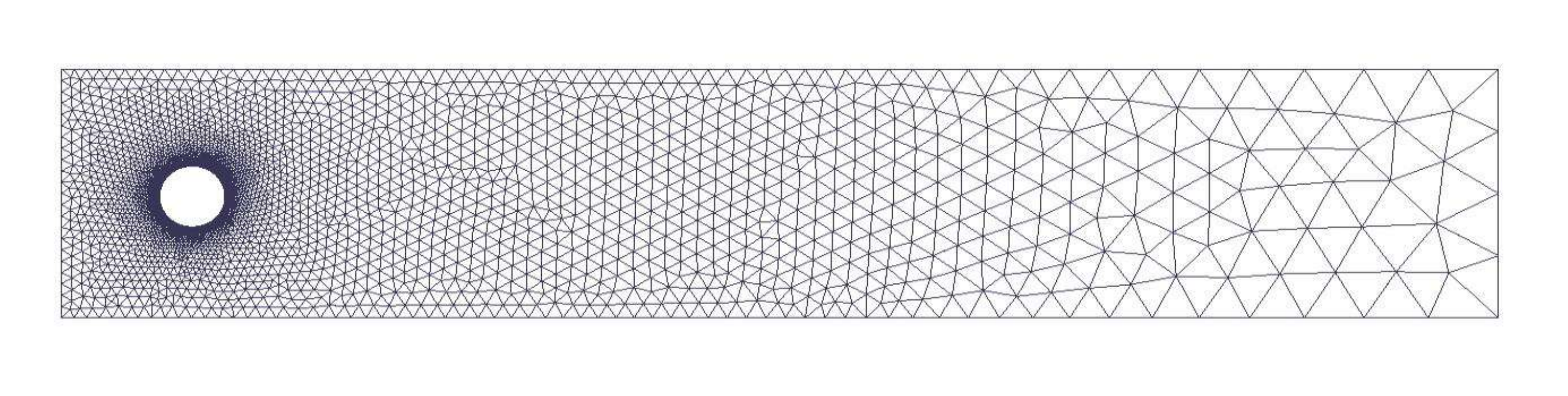}
\captionsetup{justification=centering}\caption{Unstructured triangular mesh employed in all the domain.}
\label{fig:mesh1}
\vspace{3mm}
\end{subfigure}

\begin{subfigure}[b]{\linewidth}
\centering
\includegraphics[width=0.75\linewidth]{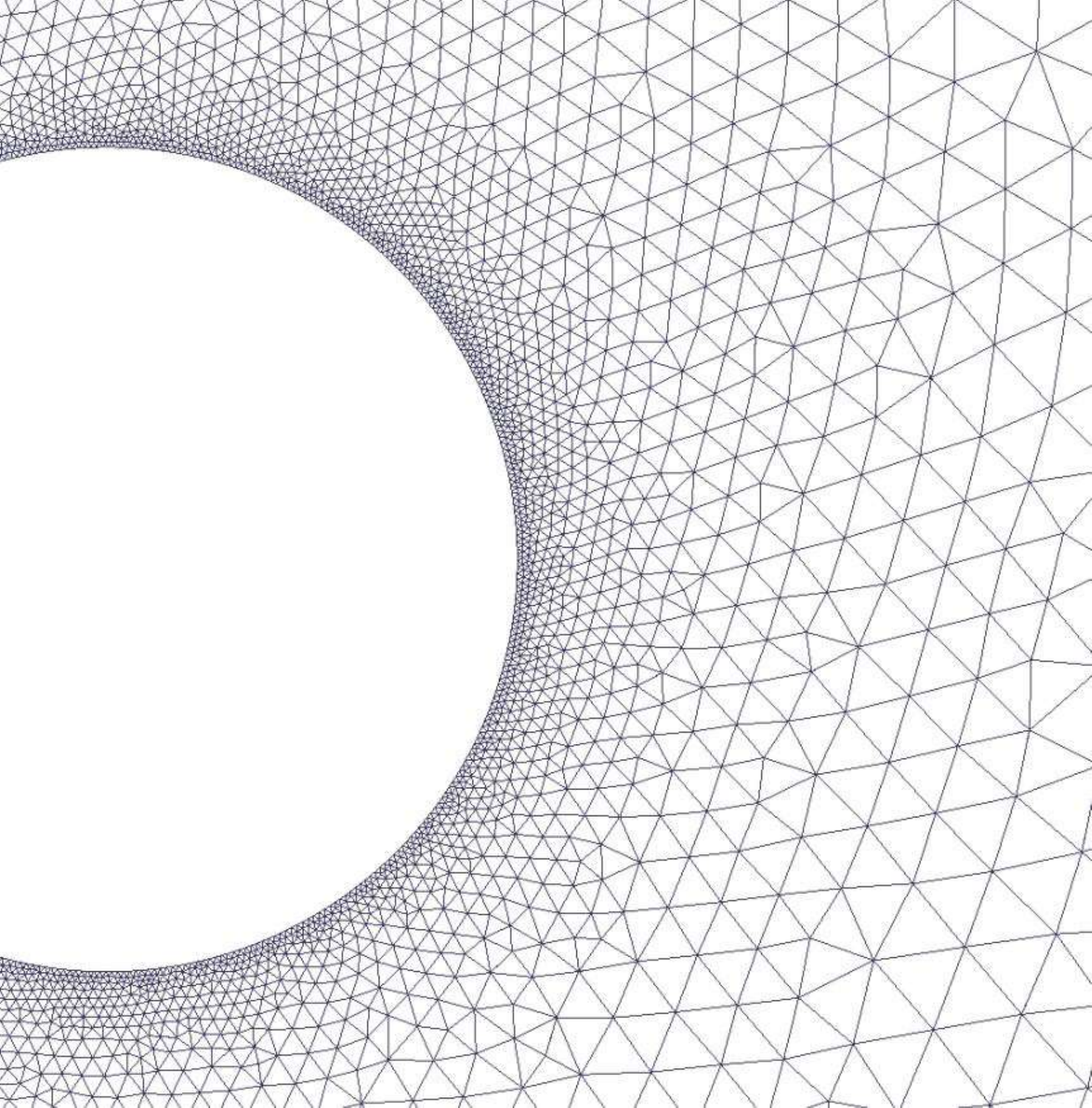}
\captionsetup{justification=centering}\caption{Detailed mesh around the cylinder.}
\label{fig:mesh2}
\vspace{3mm}
\end{subfigure}
\caption{Computational grid used for the $Re=100$ calculations.}
\label{fig:mesh}
\end{figure}

The Reynolds number $Re = \overline{U} D/\nu$ of the simulation is varied between \num{100}, \num{1000} and \num{2000}, where $\nu$ is the fluid kinematic viscosity. For the mesh, an unstructured mesh of triangular elements with refinement near the cylinder wall and the jets has been used, as shown in \autoref{fig:mesh}. The number of elements changes with the Reynolds number as expressed in \autoref{tab:params}.

\begin{table}[]
\centering
\begin{tabular}{c|ccc}
$\bm{Re}$                 & {\bfseries 100} & \textbf{1000} & \textbf{2000} \\ \hline
Mesh cells (approx.)          & 11000          & 19000           & 52000           \\
Number of witness points      & 151            & 151             & 151             \\
$|Q|_{\rm{max}}$            & 0.088        & 0.04          & 0.04          \\
$s_{\rm{norm}}$               & 1.7          & 2             & 2             \\
$C_{\rm{offset}}$             & 3.17         & 3.29          & 3.29          \\
$r_{\rm{norm}}$               & 5             & 1.25          & 1.25          \\
$w$                           & 0.2          & 1             & 1             \\
$T_k$                         & 3.37         & 3.04          & 4.39          \\
$T_a$                         & 0.25         & 0.2           & 0.2           \\
Actions per episode           & 80             & 100             & 100             \\
Number of episodes            & 350            & 1000            & 1400            \\
CPUs per environment      & 46             & 46              & 46             \\
Environments      & 1 & 1 or 20 & 20\\
Total CPUs        & 46 & 46 or 920 & 920 \\
Baseline duration             & 100          & 250           & 100           
\end{tabular}
\caption{Parameters of the simulations for each Reynolds number considered in this work.}
\label{tab:params}
\end{table}

%\begin{table}[htbp]
%  \centering
%    \begin{tabular}{r|r}
%    
%    \multicolumn{1}{c|}{\textbf{Reynolds number}} & \multicolumn{1}{c}{\textbf{Mesh cells (aprox.)}} \\
%    \midrule
%    100   & 11000 \\
%    \midrule
%    1000  & 19000 \\
%    \midrule
%    2000  & 52000 \\
% 
%    \end{tabular}%
%      \caption{Number of cells employed on the mesh for each Reynolds number considered in this work.}
%  \label{tab:params}%
%\end{table}%

Alya is used to simulate the flow. This solver assumes that the flow is viscous and incompressible, where the governing Navier--Stokes equations can be written for a domain $\Omega$ as in \autoref{eqn:NS2}:

\begin{align}
\partial_{t}\bm{u}+(\bm{u}\cdot \nabla)\bm{u}-\nabla \cdot(2\nu \bm{\epsilon})+\nabla p &= \bm{f} \qquad \textrm{in} \ \Omega  \in  (t_{0},t_{f}),\label{eqn:NS1}\\
\nabla \cdot \bm{u} &= 0 \qquad \textrm{in} \ \Omega  \in  (t_{0},t_{f}),
  \label{eqn:NS2}
\end{align}

\noindent where $\bm{\epsilon}$ is a function of the velocity $\bm{u}$ which defines the velocity strain-rate tensor ($\bm{\epsilon}=1/2(\nabla \bm{u} + \nabla^{T} \bm{u})$ ) and $\bm{f}$ are the external body forces. In \autoref{eqn:NS1}, the convective form of the nonlinear term, $\bm{C}_{\rm{nonc}}(\bm{u})=(\bm{u}\cdot\nabla)\bm{u}$, is expressed as a term conserving energy, momentum and angular momentum \cite{CHARNYI,CHARNYI2}. This form is known as EMAC (energy-, momentum- and angular-momentum-conserving equation), and its expression appears in equation \autoref{eqn:C_EMAC}. The EMAC form (\autoref{eqn:C_EMAC}) adds to the equation the conservation of energy as well as linear and angular momentum at the discrete level:

\begin{equation} 
\bm{C}_{\rm{emac}}=2\bm{u}\cdot \bm{\epsilon} +(\nabla \cdot \bm{u})\bm{u}- \frac{1}{2}\nabla |\bm{u}|^{2}.
  \label{eqn:C_EMAC}
\end{equation}

The spatial discretisation of the Navier-Stokes equations is performed by means of the finite-element method (FEM). Meanwhile, the time discretisation uses a semi-implicit Runge--Kutta scheme of second order for the convective term, and a Crank--Nicolson scheme for the diffusive term \cite{CRANKNICOLSON}. In the time integration, Alya uses an eigenvalue time-step estimation as described by \citet{LEHMKUHL2}. The complete formulation of the flow solver is described in the work by \citet{LEHMKUHL}.

At each time step, the numerical solution is obtained and the drag $F_{D}$ and lift $F_{L}$ forces are integrated over the cylinder surface $\textbf{S}$ as follows (\autoref{eqn:force}):

\begin{equation} 
\bm{F}=\int (\bm{\varsigma} \cdot \bm{n}) \cdot \bm{e}_{j} \dd{\bm{S}},
  \label{eqn:force}
\end{equation}

\noindent where $\bm{\varsigma}$ is the Cauchy stress tensor, $\bm{n}$ is the unit vector normal to the cylinder and $\bm{e}_{j}$ is a unit vector in the direction of the main flow velocity when calculating the drag and a vector perpendicular to the velocity of flow for the calculation of the lift force. The drag $C_D$ and lift $C_L$ coefficients are computed as decribed in \autoref{eqn:CD} and \autoref{eqn:CL}:

\begin{equation} 
C_{D}=\frac{2F_{D}}{\rho \overline{U}^{2}D},
  \label{eqn:CD}
\end{equation}

\begin{equation} 
C_{L}=\frac{2F_{L}}{\rho \overline{U}^{2}D}.
  \label{eqn:CL}
\end{equation}

\subsection{DRL setup}\label{sec:DRL_setup}

\begin{figure}[H]
\includegraphics[width=\linewidth]{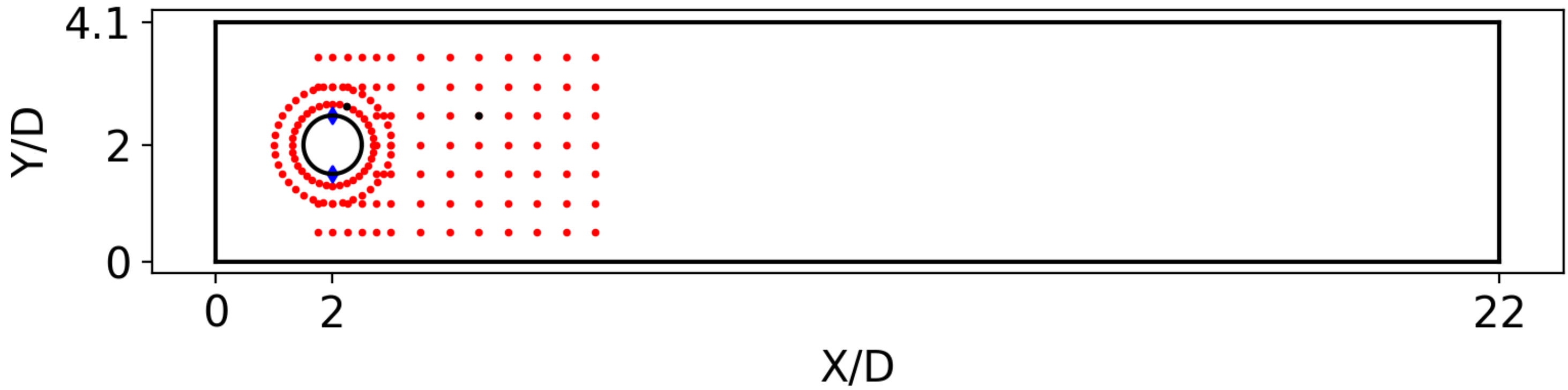}
\caption{Schematic representation of the computational domain, where the red dots correspond to the location of the probes. The position of two probes is remarked as black dots for further analysis in \autoref{cross-t}.}
\label{fig:witness}
\end{figure}

As discussed in the introduction, the DRL interacts with the domain through three channels. The first channel is the observation state ($s$), based on the extraction of pressure values at a series of predefined points along the domain. These points, known as witness points or probes, are located in the same positions as in Rabault \textit{et al.} \cite{RABAULT1}. There are \num{151} witness points in total distributed around the cylinder and along the wake as shown in \autoref{fig:witness}. The values of the pressure obtained at the witness points are normalised by a factor $s_{\rm{norm}}$ so that the state values given to the agent are between $-1$ and $1$, approximately. The values of $s_{\rm{norm}}$ for each Reynolds-number case are given in \autoref{tab:params}.

The second channel of interaction between the DRL and the numerical simulation is the action that is given by the control of the jets on the cylinder ($a$). The value of the action $a$ is directly related with the control value of the upper jet intensity $Q_1$, while the bottom jet will do the opposite control to ensure a global zero mass flow rate between both jets, i.e. $Q_{2}=-Q_{1}$. This is a more realistic control and helps to make the numerical scheme more stable as reflected by \citet{RABAULT1}. During the training, the maximum value of $|Q_1|$ is limited to $|Q_1| < 0.06\;Q_{\rm{ref}} \approx 0.88$ for the $Re = 100$ case, as in the work by \citet{RABAULT1}, to avoid unrealistically large actuations. Note that $Q_{\rm{ref}}$ is the mass flow rate intercepting the cylinder, and it is calculated as in \autoref{eqn:Q_ref}:

\begin{equation}
Q_\mathrm{ref} = \int_{-D/2}^{D/2}\rho U_\textnormal{in}(y) \dd{y}.
\label{eqn:Q_ref}
\end{equation}

\begin{figure}[H]
\centering
\includegraphics[width=0.75\textwidth]{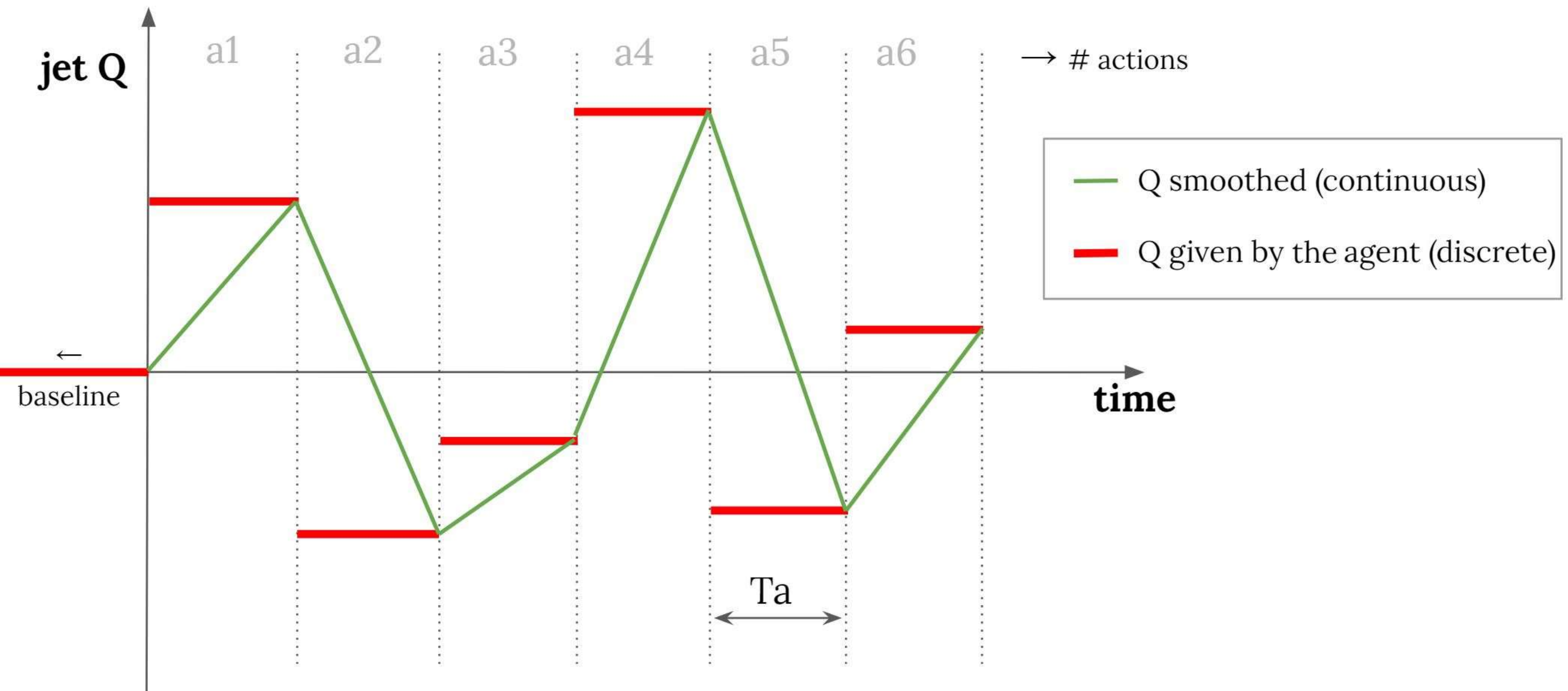}
\caption{Flow rate of the jet $Q$ smoothed and applied in the numerical simulation (green) versus the discrete $Q$ decided by the DRL agent (red).}
\label{fig:smoocroquis}
\end{figure}

For higher Reynolds numbers this clipping value of $|Q_1|$ is reduced to $0.04$. In addition, even though the value of $Q$ selected by the ANN is unique in each action, $Q$ is not imposed as a constant value during the entire action duration. In particular, $Q$ is assessed as a curve to prevent significant changes in the boundary condition between actions that could lead to numerical discrepancies, similar to how the smooth control was presented by \citet{RABAULT3}. This way, the imposed mass flow starts from the previous value $Q_{0}$ and increases or decreases linearly until the new value $Q_{1}$ has been achieved during the entire action time $T_{a}=t_{1}-t_{0}$. Consequently, $Q$ in \autoref{eqn:jet_A} is presented as a function of time in \autoref{eqn:Q_t} and is illustrated in \autoref{fig:smoocroquis}.

\begin{equation} 
Q(t)=\frac{Q_{1}-Q_{0}}{T_{a}}(t-t_{0})+Q_{0}.
  \label{eqn:Q_t}
\end{equation}

It should be noted that the first action of each episode (an episode is understood as a sequence of interactions between the neural network and the simulation, which generates the input data for the agent algorithm) always starts from a baseline case, \textit{i.e.} a periodically stable flow without jet actuation. 
This baseline case uses the same domain, and the flow is fully developed without applying jet control.

The third interaction channel between the DRL and the numerical simulation is the reward or goal ($r$), which in this case is aimed at minimising the cylinder drag and it is defined as in \autoref{eqn:reward}:

\begin{equation} 
r=r_{\rm{norm}}(-\langle \overline{C_{D}}\rangle - w \langle \overline{C_{L}} \rangle + C_\textnormal{offset}),
  \label{eqn:reward}
\end{equation}

\noindent where $\langle\cdot\rangle$ indicates averaging over a baseline vortex-shedding period, $w$ is a lift-penalisation factor, $C_{\rm{offset}}$ is a coefficient to centre the initial reward around $0$, obtained from the value of $r$ at the end of the baseline simulation, and $r_{\rm{norm}}$ is used to normalise this reward between $0$ and $1$ approximately. In such a way, the agent receives the reward in an optimal range. Note that the value of $r_{\rm{norm}}$ is set from an a priori guess of the expected the maximum reward. The lift-penalisation factor is set in such a way that the drag is minimised, and at the same time the effect of the possible growth of induced lift is mitigated. If this lift penalisation is not introduced into the reward function, the agent can find a strategy where both jets blow in the same direction at their maximum strength, as discussed in \citet{RABAULT2}. The values of $w$, $C_{\rm{offset}}$ and $r_{\rm{norm}}$ are given in \autoref{tab:params} for the different Reynolds numbers.

In a nutshell, the agent will choose an action ($a$) given a specific state ($s$) in order to maximise a reward ($r$). The function that determines what will be the reward is a normal distribution. During the training process, the agent will choose an action around the average of this normal distribution. This is known as the exploration noise and it helps the method to converge towards a better solution. Once the training has been performed, the agent can be tested in a deterministic mode. Here, the most probable action in the normal distribution is chosen in order to maximise the reward with the learning obtained during the training.

Following the work by \citet{RABAULT1}, the ANN is designed with two dense layers of 512 neurons and a PPO agent is selected to carry out the control. The PPO agent follows a policy-gradient method in order to obtain the weights of the ANN. The implementation of the DRL is done through the open-source Tensorforce library \cite{TENSORFORCE}, which is built from the TensorFlow open-source library \cite{TENSORFLOW}, and it includes defining and creating both the ANN and the control agent. The selection of the initial parameters of the DRL depends intrinsically on the problem itself. In this case, the total number of actions is related to the vortex-shedding period $T_{k}=1/f_{k}$ through the Strouhal number $St=f_{k} \cdot D/\overline{U}$.

\begin{figure}[H]
\begin{subfigure}[b]{0.48\linewidth}
\includegraphics[width=\linewidth]{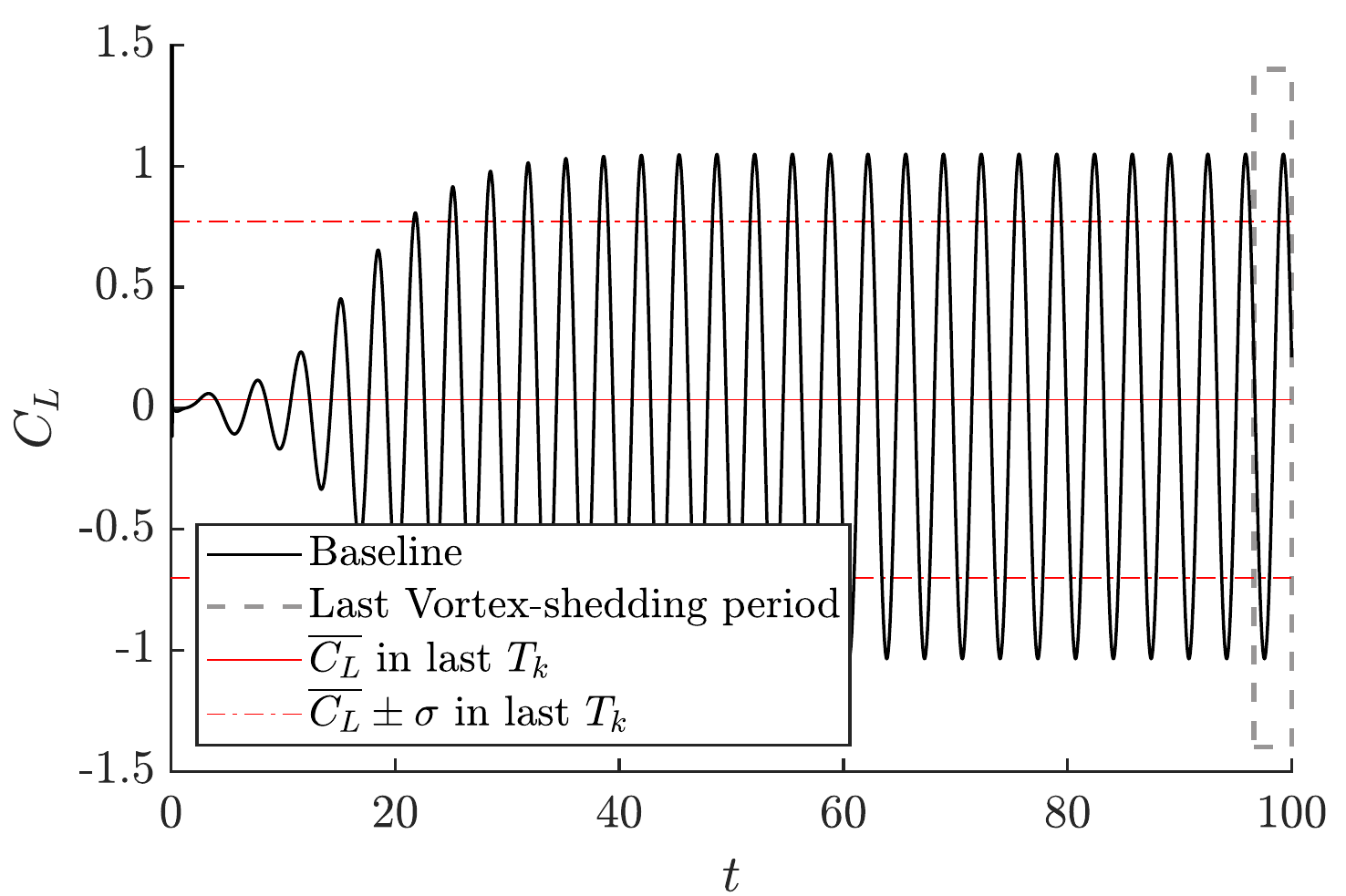}
\captionsetup{justification=centering}
\label{fig:BS_CL}
\end{subfigure}
\begin{subfigure}[b]{0.48\linewidth}
\includegraphics[width=\linewidth]{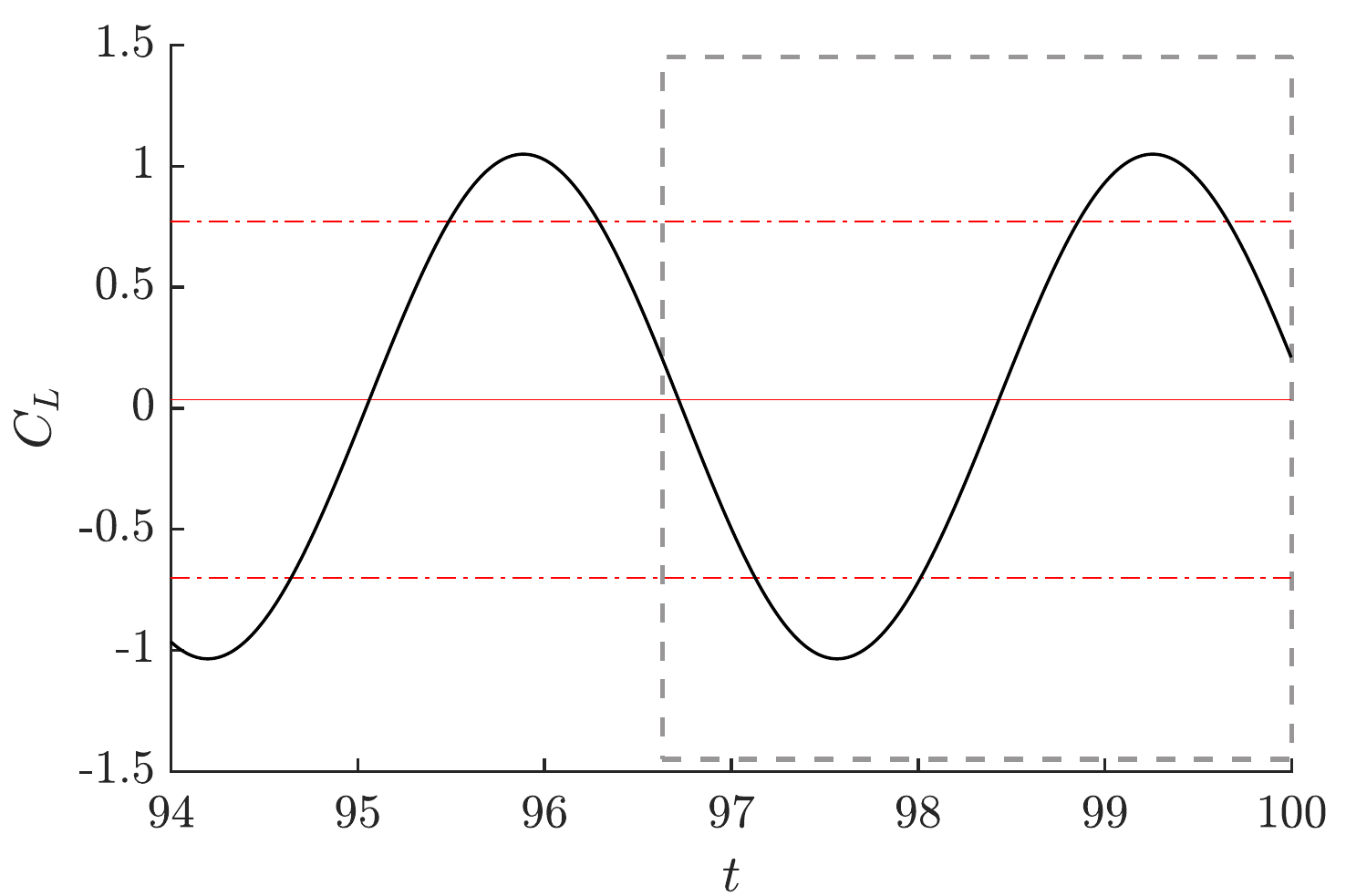}
\captionsetup{justification=centering}
\label{fig:BS_CL_ZOOM}
\end{subfigure}
\caption{Temporal evolution of the lift coefficient in the baseline case (without jet actuation). (a) Baseline lift-coefficient results. The vortex-shedding period is $3.37$ time units. (b) Zoom-in view of the vortex-shedding period in the baseline case.}
\label{fig:BS_CL_SUB}
\end{figure}

\autoref{fig:BS_CL_SUB} shows the lift coefficient of the baseline case. This figure shows that the vortex-shedding period is $T_{k}=3.37$ time units for the $Re = 100$ case and $St=0.29$; this is in agreement with both experiments \cite{SCHAFER} and simulations \cite{RABAULT1}. Taking the vortex-shedding period as reference, the action time $T_a$ is defined as \SI{7.5}{\percent} of $T_k$, as in Rabault \textit{et al.} \cite{RABAULT1}. This period of actuation was found to be large enough so the consequences of the actuation can be perceived by the flow, and small enough so that the actuations can anticipate and adapt to the needs of the control. In the higher-$Re$ cases, this actuation period is reduced to $T_a = 0.2$ due to their more chaotic flow behaviour and, therefore, the necessity to adapt the actuations faster. Based on this simulation, it can be concluded that before starting the DRL, the baseline case needs to be run for over $50$ time units in order to obtain a periodic stabilised flow. As suggested in the work of \citet{RABAULT1}, at $Re = 100$ the typical time of an episode should be between 6 and 8 vortex sheddings so that the agent has enough time to learn the suitable control policy. A total of $80$ actuations are carried out on each episode, resulting in an episode duration of $20$ time units. In the case of higher Reynolds number, a total of $100$ actions are conducted in each episode, since the duration of each actuation is reduced. 

\begin{figure}[H]
\centering
\includegraphics[width=0.75\textwidth]{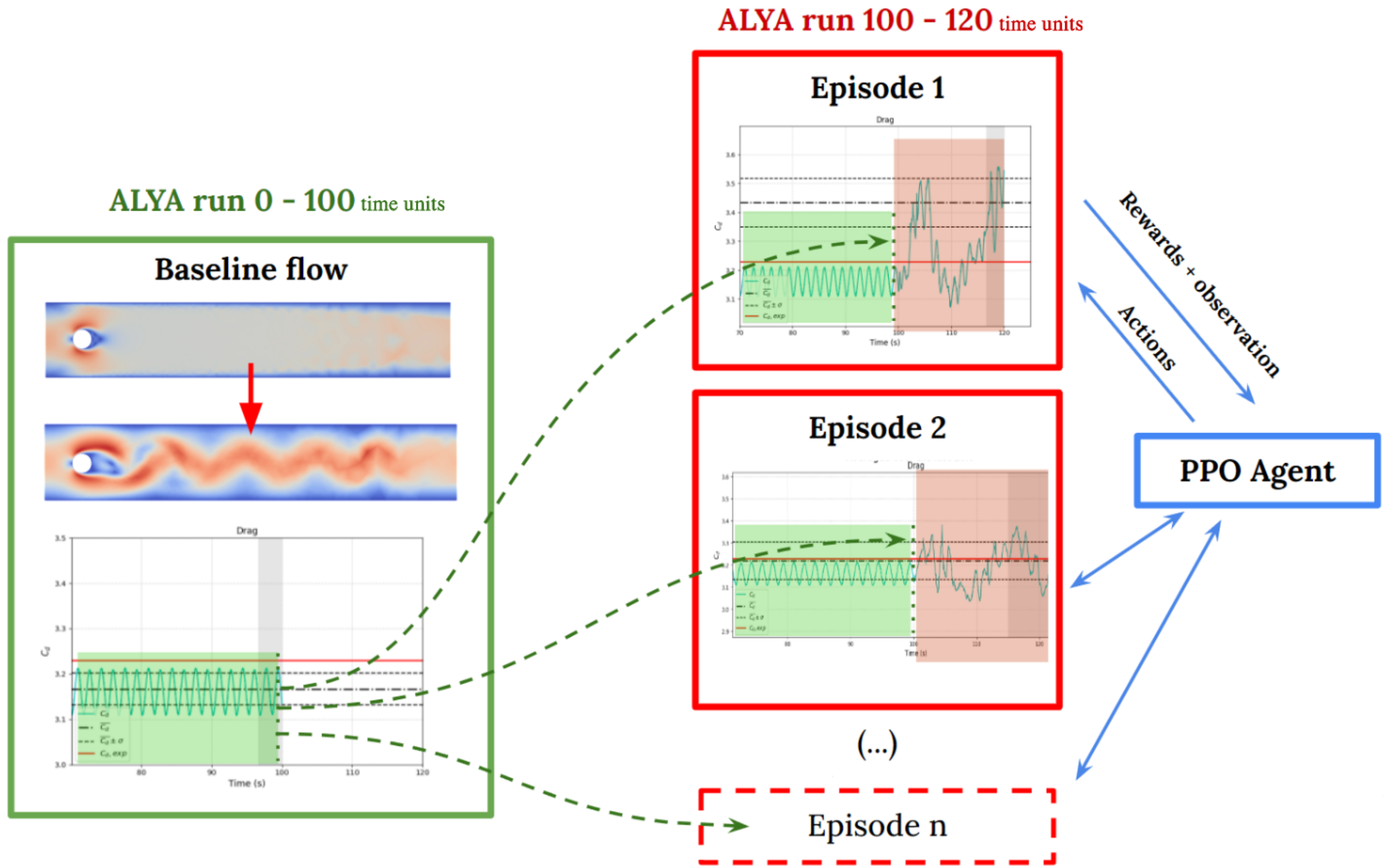}
\caption{General overview of the DRL-CFD (computational fluid dynamics) framework employed in this work. A multi-environment approach is used to parallelise the learning and to speed up training.}
\label{fig:generalview}
\end{figure}

For the cases at high Reynolds number of $1000$ and $2000$ parallel environment framework has been adopted to speed up the learning process using a total of 20 environments. Also, in the case at $Re = 1000$ a comparison of the results is done when a single environment is employed. A total of 46 Marenostrum IV central-processing units (CPUs) are used on each environment. Therefore, a total of 46 or 920 CPUs are used depending on whether a single environment or 20 environments are considered on the case, respectively. All these parameters are collected in \autoref{tab:params}. The general DRL framework is summarised in \autoref{fig:generalview}.

\section{Results and discussion}\label{sec:Results}

This section is divided into three parts. First, the results obtained using the Alya solver and the DRL applied are validated using literature data. Once validated, higher Reynolds number results are obtained and discussed, for which no precedent has been found in the literature. Finally, a cross-application of agents is analysed to save computational resources in resolving cases with high Reynolds numbers.

\subsection{CFD and DRL code validation}

As can be seen in the work by \citet{RABAULT1} for a $Re=100$, through the correct actuation of the DRL, it was possible to achieve a decrease in cylinder drag of \SI{8}{\percent}. These results have been validated in other simulations, such as \citet{LI}, and applied in similar research \cite{TOKAREV,ELHAWARY,HAN}.% The baseline simulation is shown in \autoref{fig:BS_CD}. These values are optimised and are listed in \autoref{tbl:table_setup}. (XX MOVE TABLE TO THE END WITH MORE REYNOLDS AND SUPPRESS COMMENT?) 
%\begin{figure}[H]
%	\includegraphics[width=0.95\linewidth]{CD_baseline.pdf}
%	\caption{Drag coefficient of the baseline case, where the average $C_{D}$ in the last vortex shedding is 3.17. Dashed gray shows the period corresponding to the last shedding vortex. Solid red shows the mean of the coefficient during the last vortex shedding, while dash-dotted red shows the standard deviation in this period.}
%	\label{fig:BS_CD}
%\end{figure}

\begin{figure}[H]
	\includegraphics[width=0.95\linewidth]{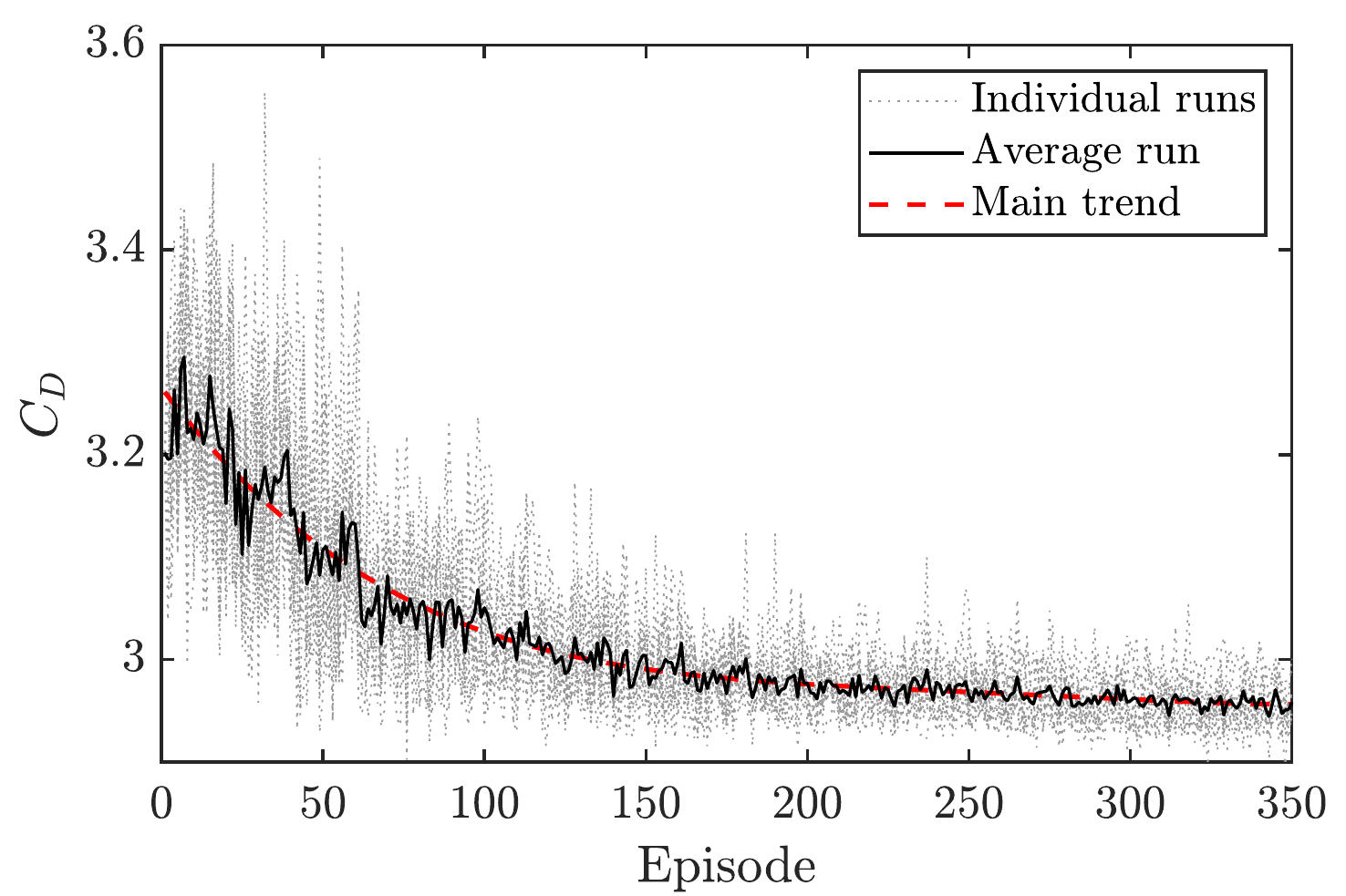}
	\caption{Evolution of $C_D$ in the last vortex-shedding period as a function of the episode.}
	\label{fig:allcd}
\end{figure}

\autoref{fig:allcd} shows eleven different trainings launched and the average $C_{D}$ obtained for the last vortex-shedding period of the different episodes. It can be seen that all cases follow the same learning trend, where most of the said learning is observed in the first hundred episodes. The main trend is obtained by adjusting a fourth-degree polynomial fit using the average coefficient data, and its purpose is to help to visualise the data. From this point on, the slope of the $C_{D}$ decrease is not so pronounced, reaching a stable solution for 350 episodes.

\begin{figure}[H]
\begin{subfigure}[b]{0.48\linewidth}
\includegraphics[width=\linewidth]{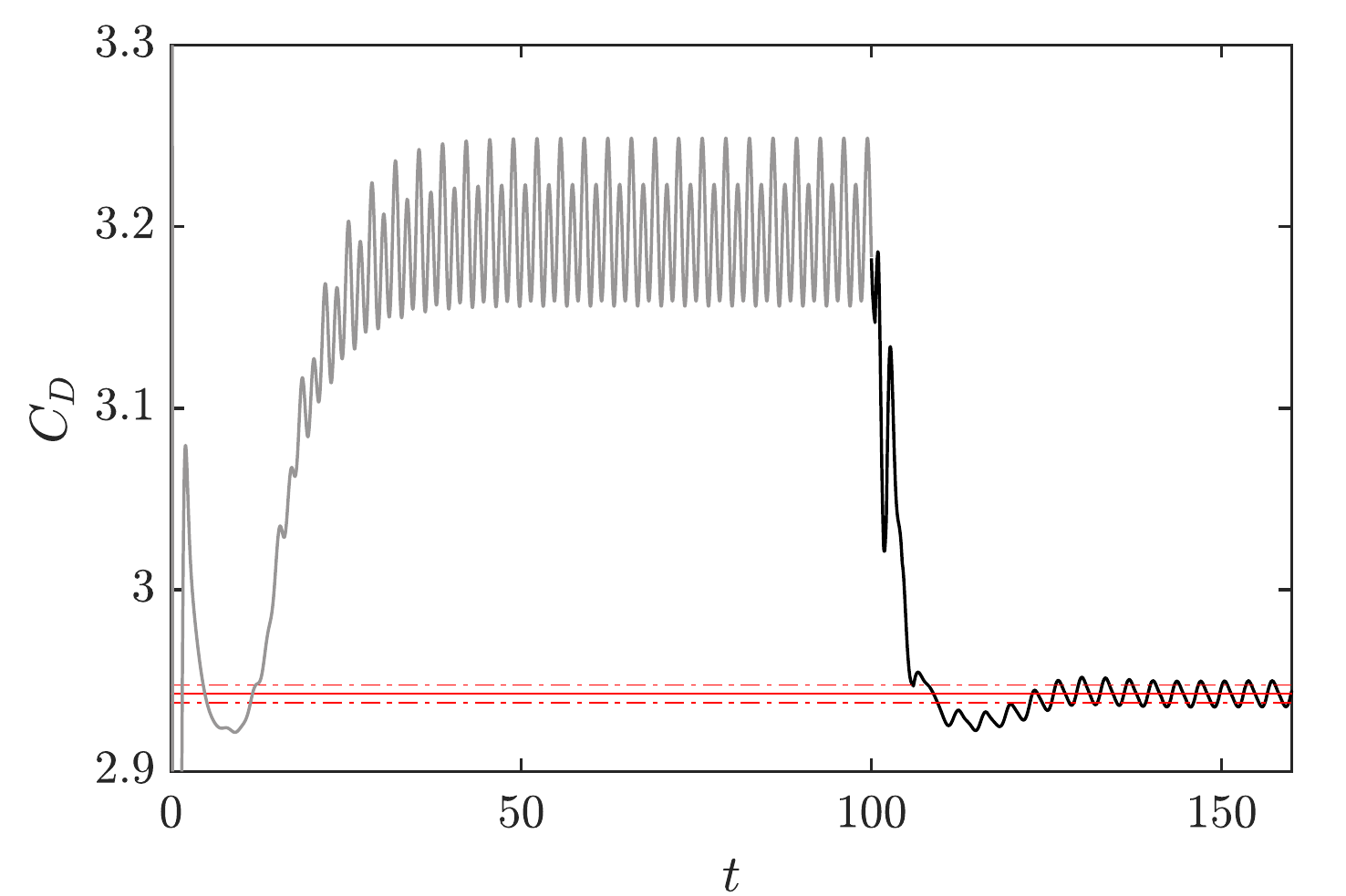}
\captionsetup{justification=centering}
\label{fig:cd_sr}
\end{subfigure}
\begin{subfigure}[b]{0.48\linewidth}
\includegraphics[width=\linewidth]{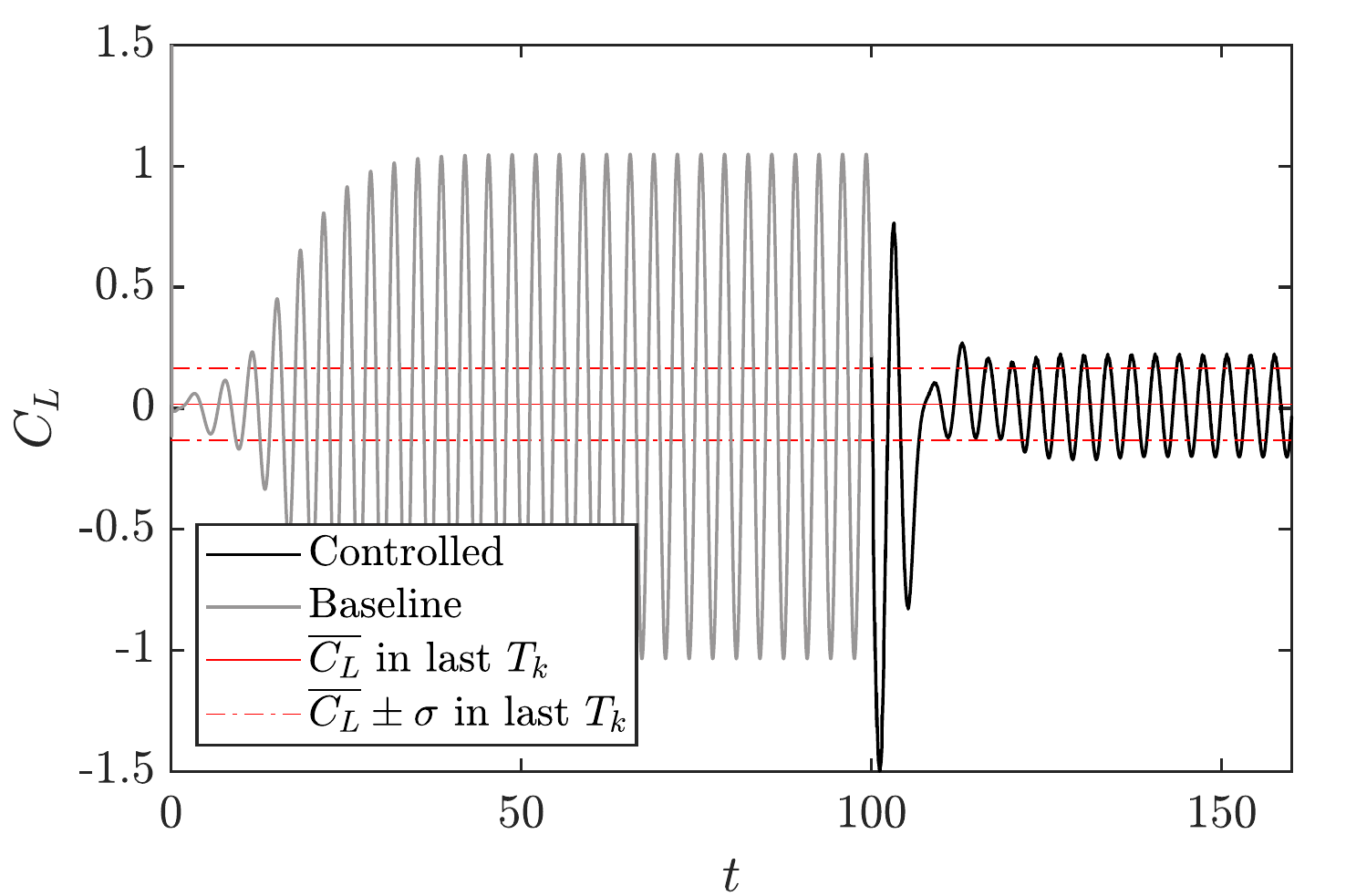}
\captionsetup{justification=centering}
\label{fig:cl_sr}
\end{subfigure}
\caption{Temporal evolution of $C_D$ (left) and $C_L$ (right) obtained through the application of the DRL control (at $t=100$), run in a deterministic mode, for $Re = 100$.}
\label{fig:cdcl_sr}
\end{figure}

The obtained learning is then applied to the previously-shown baseline. To this end, one of the $11$ trained agents is ranromly selected and is run in a deterministic mode. This deterministic simulation is initiated from the conditions of the baseline case at $100$ time units from the start. The $C_{D}$ and $C_{L}$ trends are shown in \autoref{fig:cdcl_sr}, which indicates that a  value of $C_D=2.95$ is obtained after applying the control, {\it i.e.} a decrease of \SI{8.9}{\percent} compared with the baseline case. The $C_D$ improvement is calculated as a function of the baseline $C_{D}$:

\begin{equation}
\textnormal{improvement} = \left[1 - \dfrac{{C_D}^\textnormal{controlled}}{{C_D}^\textnormal{baseline}}\right]\times 100.
\label{eqn:cd_improvement}
\end{equation}

This result is rapidly achieved from the moment the control is applied. In the case of $C_{L}$, the mean remains approximately at \num{0}. However, the amplitude of the variation of this coefficient has been reduced, as shown by the lower standard deviation obtained. This may indeed have great beneficial consequences from the structural and stability points of view.

\begin{figure}[H]
\centering
	\includegraphics[width=0.6\linewidth]{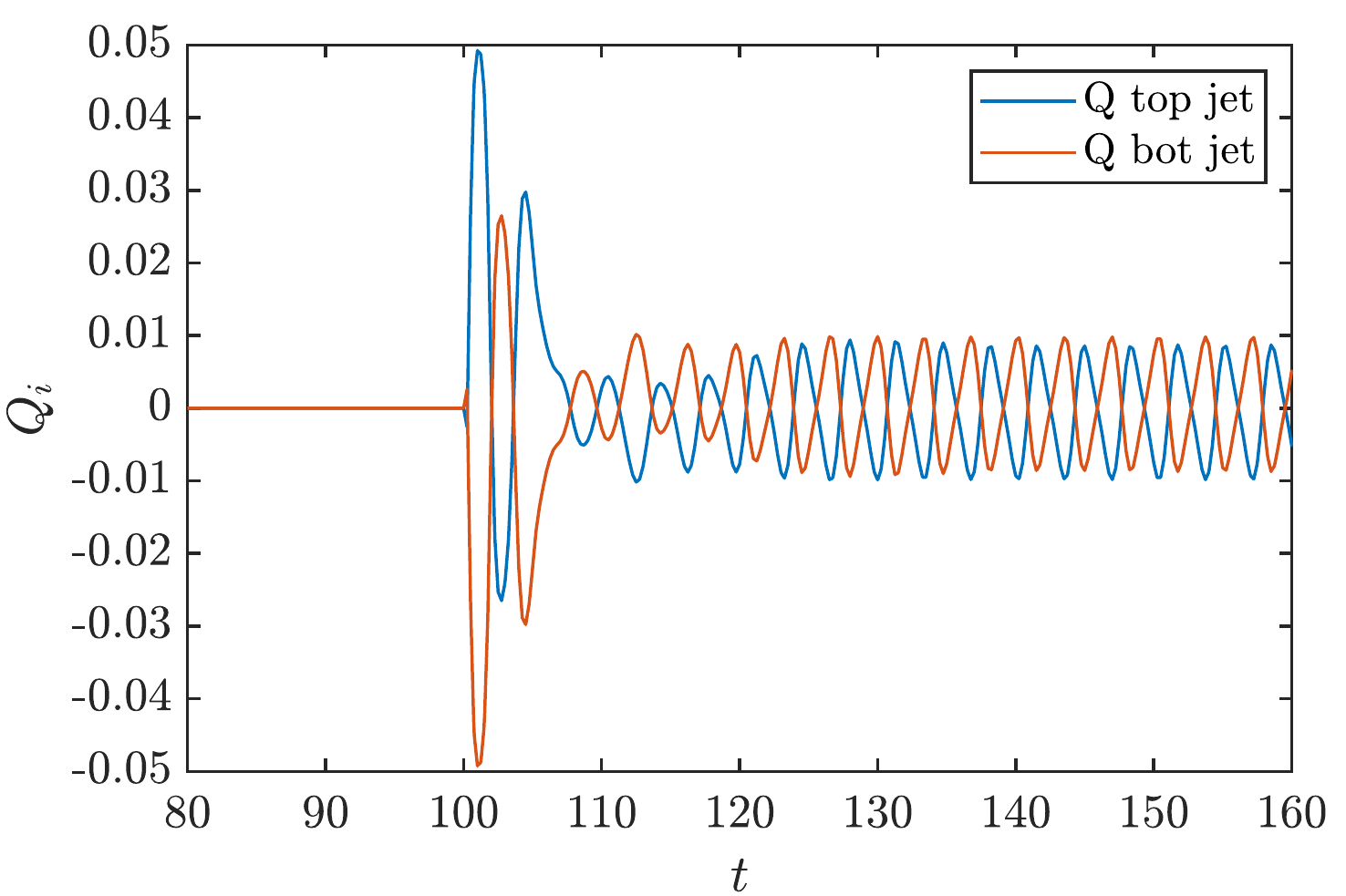}
	\caption{Flow rate through each jet as a function of time after applying the DRL control.}
	\label{fig:q_sr}
\end{figure}

The control imposed on each jet is represented in \autoref{fig:q_sr}. As discussed above, the control starts at \num{100} time units of simulation, so the values of injected or suctioned mass prior to this time are null. Having imposed the synthetic-jet condition, everything injected by one jet (positive values) will be equivalent to what is suctioned by the other (negative values). In the first \num{10} time units approximately, the control of greater amplitude leads to a significant reduction in the drag, maximising the reward. Next, a transitional control seeks to obtain a more stable actuation. After \num{25} time units, the solution is practically periodic.

\begin{figure}[H]
\centering
\subfloat[Instantaneous velocity-magnitude fields.]{\includegraphics[height=2.8in]{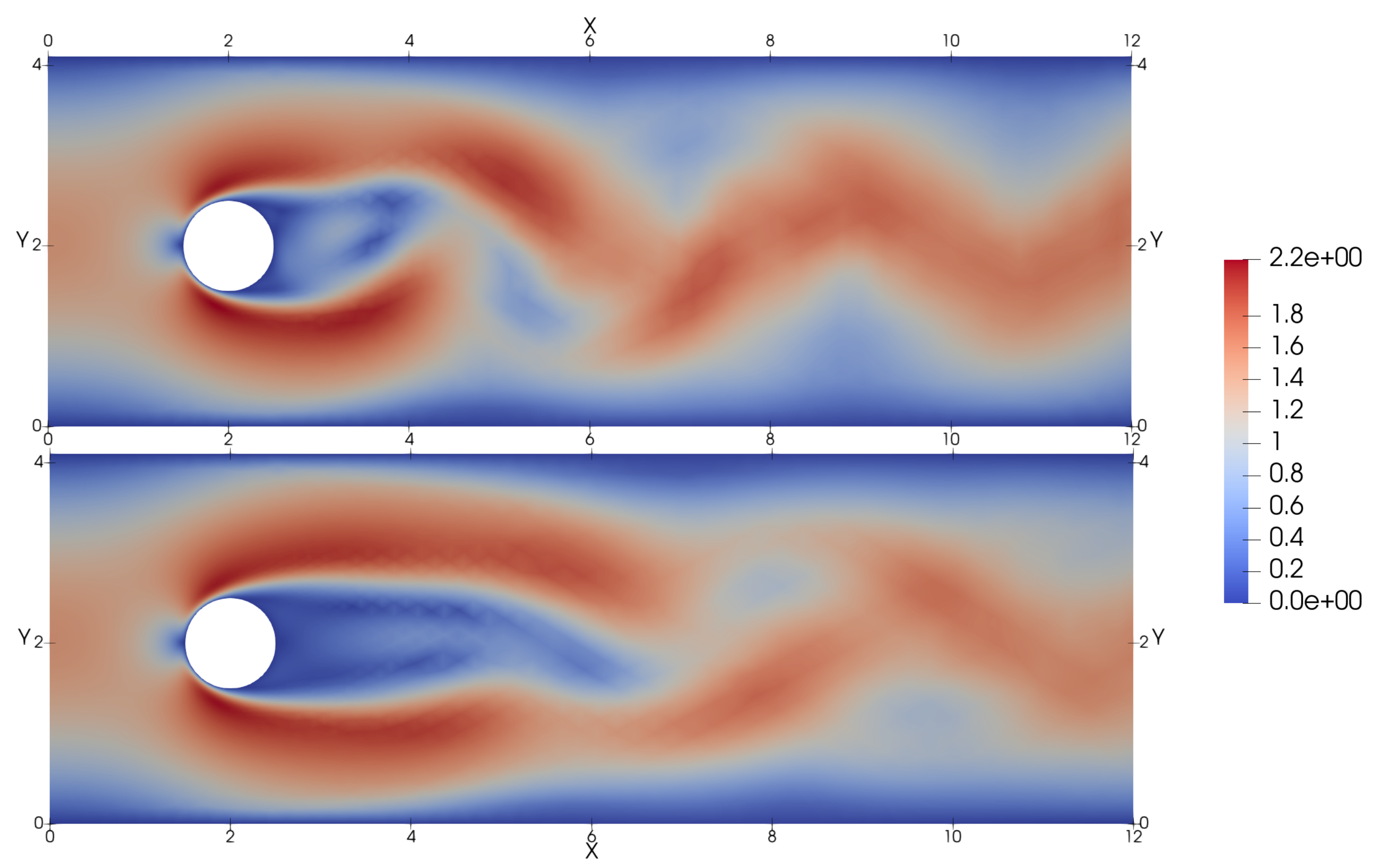}}\\
\vspace{5pt}
\subfloat[Instantaneous pressure fields.]{\includegraphics[height=2.8in]{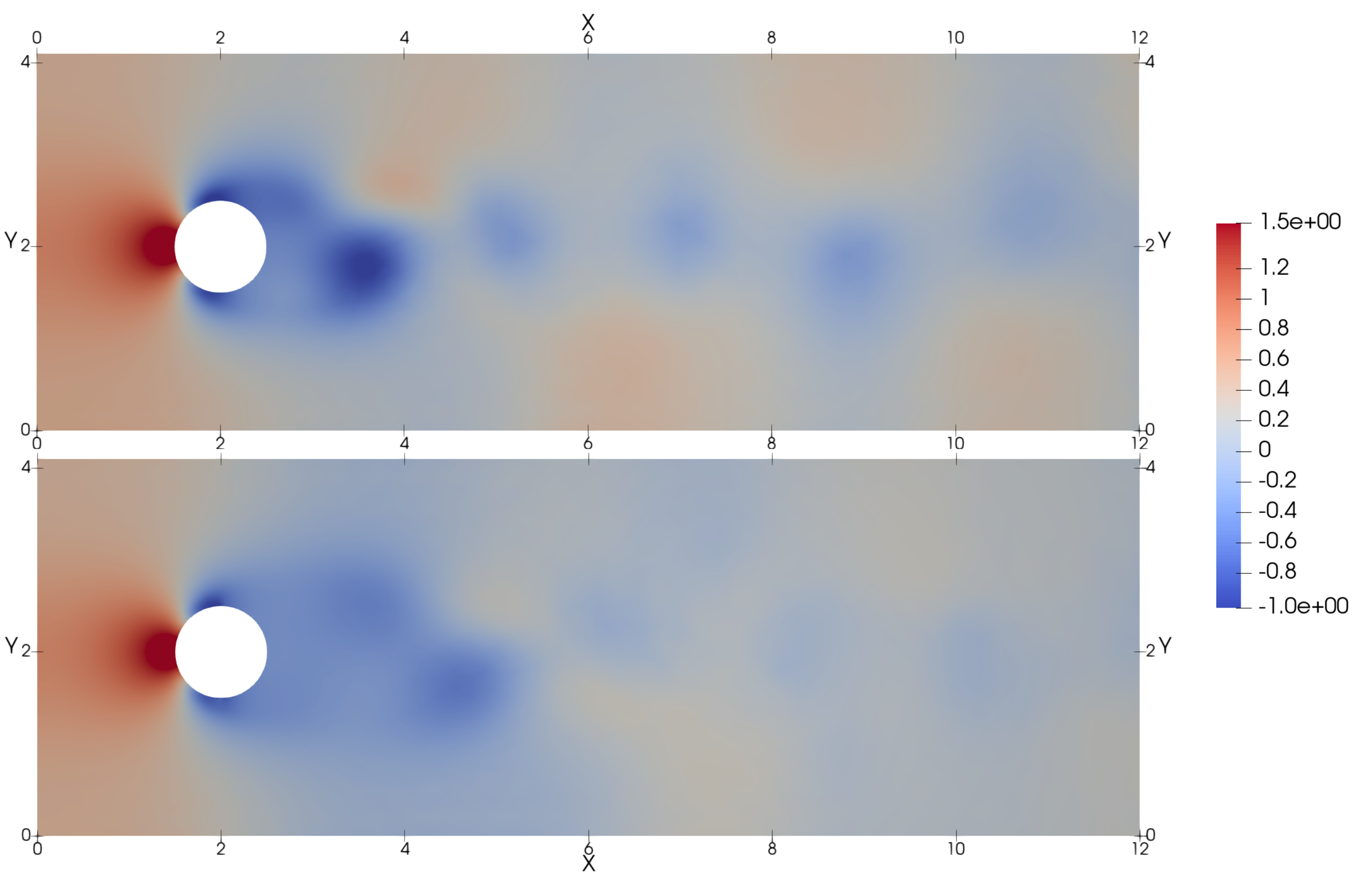}}
\caption{Instantaneous flow fields at $Re=100$, where for each pair of images, the baseline case without control is depicted on the top and the controlled case is depicted on the bottom.}
\label{fig:VelyPcontrol}
\end{figure}

More information about the control is obtained by observing the contours of instantaneous velocity and pressure for the controlled and uncontrolled cases, as shown in \autoref{fig:VelyPcontrol}. The DRL agent reduces $C_{D}$ by manipulating the wake vortex: it increases the size of the recirculation region and reduces both the frequency and the amplitude of the von Kármán street downstream of the cylinder. A similar conclusion is obtained from the instantaneous pressure field, which exhibit a decrease in the pressure maxima after applying the control. Comparing our results with those by \citet{RABAULT1}, it can be seen that the response of the control is practically the same except for very small differences that can come from the different numerical schemes to resolve the flow. %These differences are mainly due to changes in the mesh between the two cases and to the fact that the Alya solver is more dissipative than the FENICS solver used in \citet{RABAULT1}.

Increasing the Reynolds number also requires reducing the period of actuation, as mentioned in \autoref{sec:DRL_setup}. Additionally, increasing the Reynolds number leads to a more complex flow, thus a higher number of episodes are expected to be necessary to complete the training. Therefore, parallelisation using 20 environments has been adopted when simulating higher-$Re$ cases. The parallelisation using a multi-environment approach is fully detailed in the work of \citet{RABAULT2}.

\begin{figure}[H]
\begin{subfigure}[b]{0.48\linewidth}
\includegraphics[width=\linewidth]{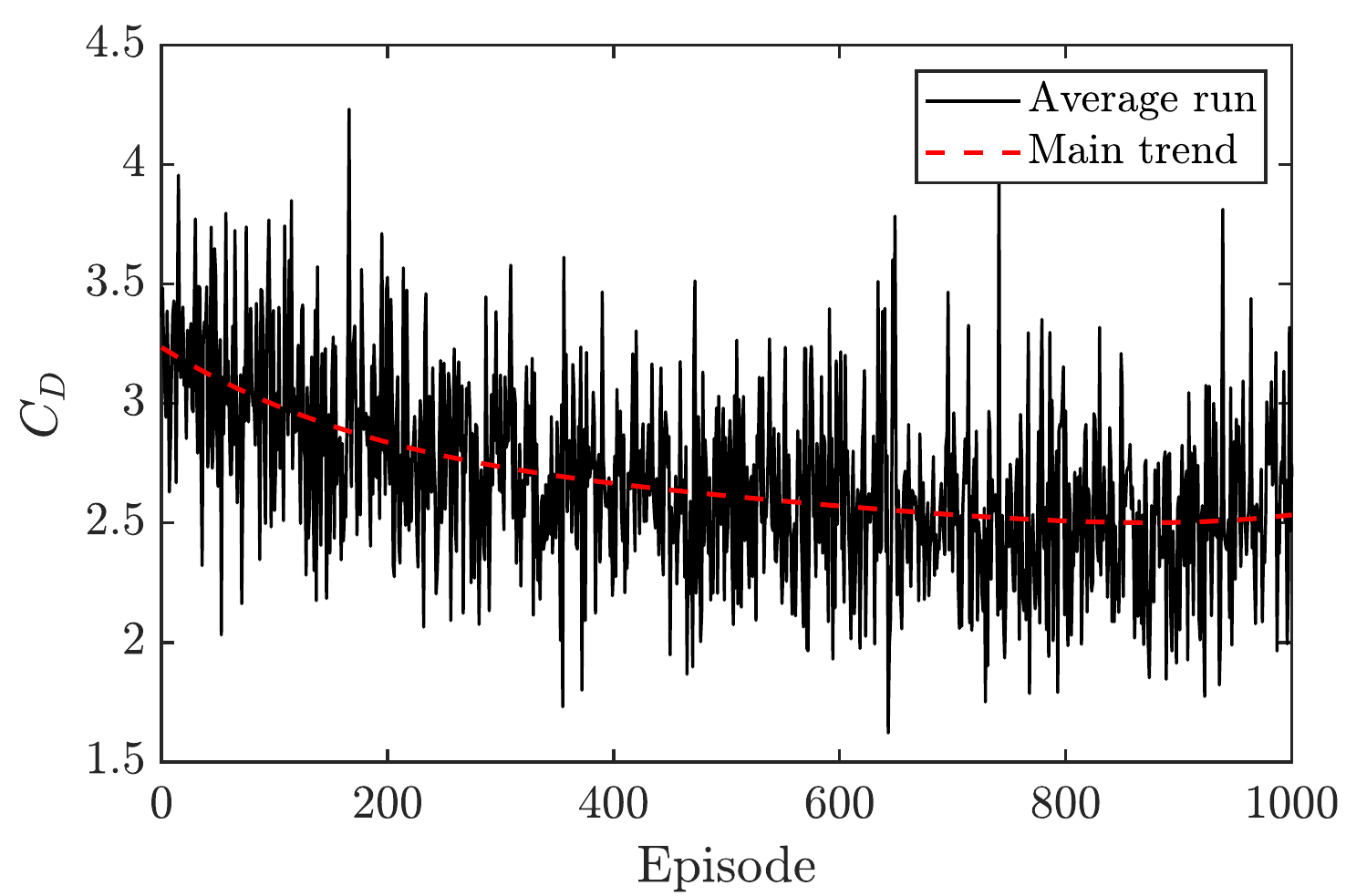}
\captionsetup{justification=centering}
\label{fig:cd_single1000}
\end{subfigure}
\begin{subfigure}[b]{0.48\linewidth}
\includegraphics[width=\linewidth]{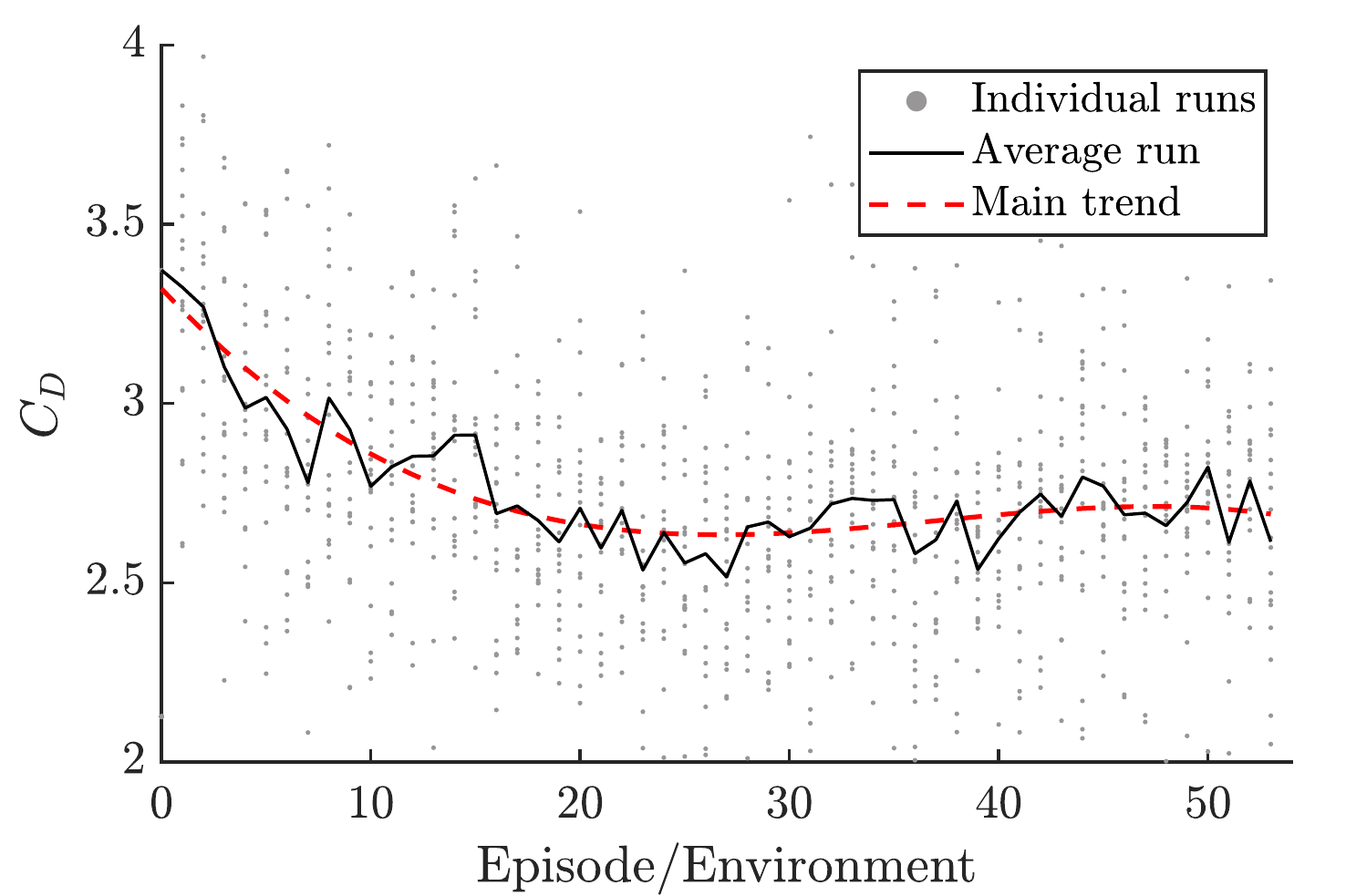}
\captionsetup{justification=centering}
\label{fig:cd_multi1000}
\end{subfigure}
\caption{Evolution of $C_D$ in the last vortex-shedding for each episode employing a single environment (left) or a 20-multi-environment approach (right) showing the learning curve indexed by the episode number for one of the 20 environments.}
\label{fig:cd1000}
\end{figure}

In \autoref{fig:cd1000}, a comparison between employing a single environment and a 20 multi-environment approach is conducted for a case of $Re=1000$. It can be observed that if the Reynolds increases an order of magnitude, in this case, the number of episodes for the DRL to act to the same degree is three times greater using only one environment, as compared with \autoref{fig:allcd}. The same solution can be obtained by parallelising through 20 environments, using 50 episodes in each environment, which helps to reduce the real calculation time significantly.

\begin{figure}[H]
\centering
\subfloat[Instantaneous velocity-magnitude fields.]{\includegraphics[height=2.85in]{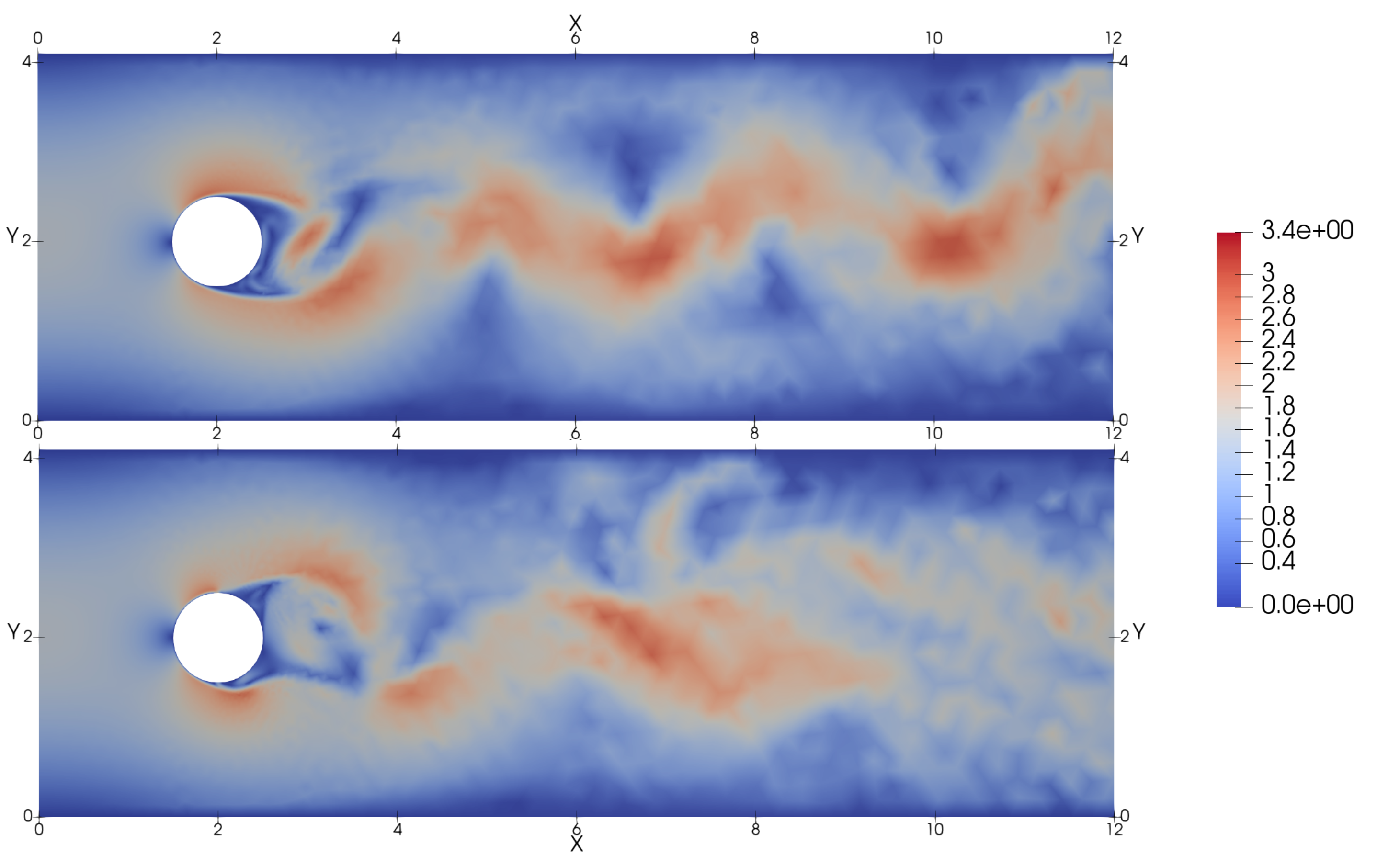}}\\
\vspace{3pt}
\subfloat[Average velocity-magnitude fields, where the mean streamlines are indicated in black.]{\includegraphics[height=2.8in]{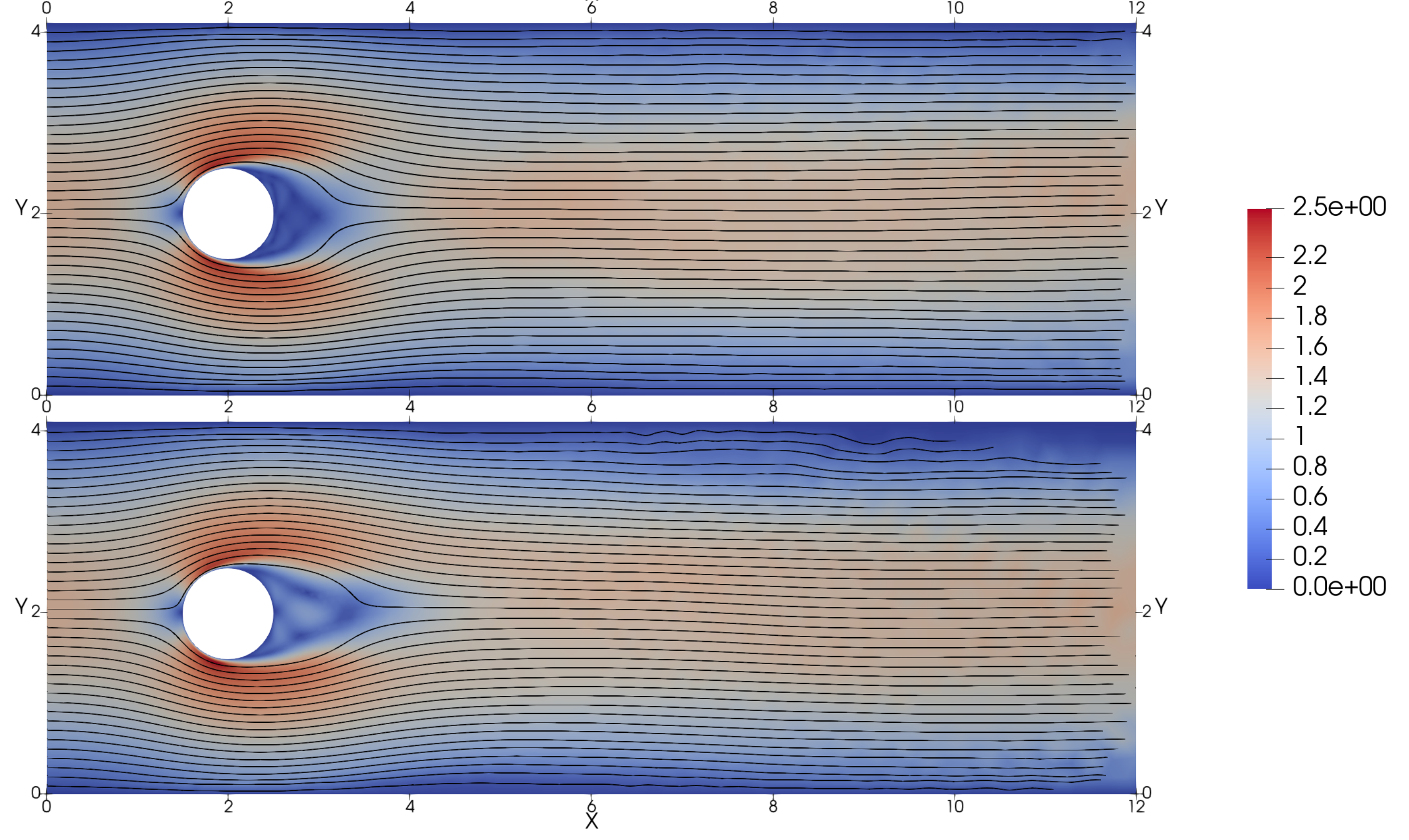}}
\caption{Instantaneous and average flow fields at $Re=1000$, where for each pair of images, the baseline case without control is depicted on the top and the controlled case is depicted on the bottom.}
\label{fig:Vel1000}
\end{figure}

As mentioned before, the flow is more complex, and its structures are less periodic and more chaotic. To analyse the flow, the instantaneous and average velocity field are represented in \autoref{fig:Vel1000}.

The obtained results are consistent with those of \citet{REN}, which reports an extensive flow analysis of the same Reynolds number. The control produces an elongation of the recirculation bubble behind the cylinder, reducing at the same time the hydrodynamic drag. For this Reynolds number, the control strategy found by the PPO agent is based on synchronous blowing. This strategy is similar to that found in lower Reynolds numbers: the jets produce an ejection or suction, generating a flow opposite to that caused by the wake.

\subsection{DRL application at Reynolds number 2000}\label{DRL_2000}

Through the use of parallelisation with a multi-environment strategy, even higher Reynolds numbers can be achieved. The results obtained for a training with $Re = 2000$ are discussed below and, to the authors' knowledge, there is no record of applying DRL control to such a high Reynolds number in the literature.

\begin{figure}[H]
\begin{subfigure}[b]{0.48\linewidth}
\includegraphics[width=\linewidth]{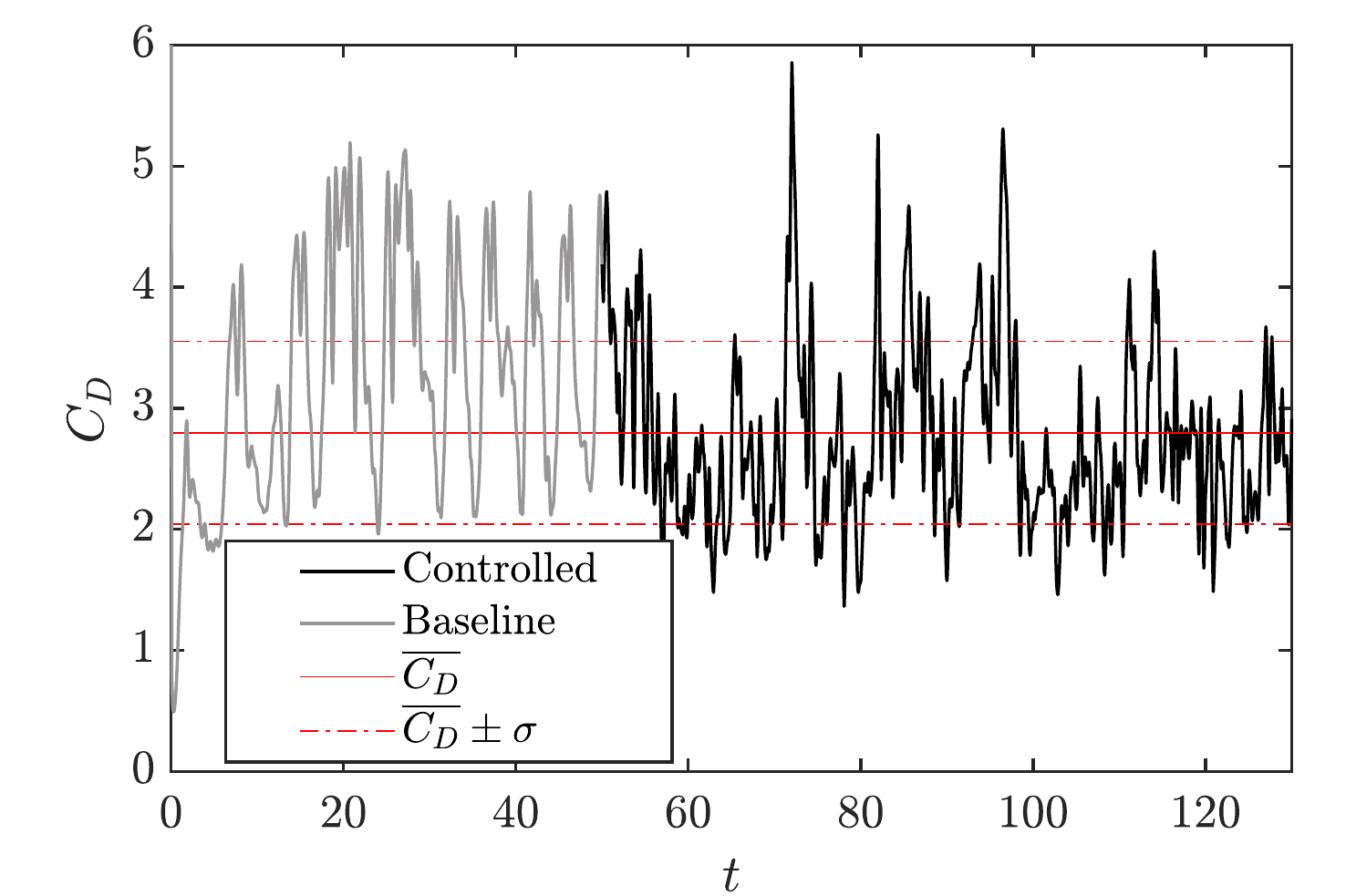}
\captionsetup{justification=centering}
\label{fig:cd_sr_2t2}
\end{subfigure}
\begin{subfigure}[b]{0.48\linewidth}
\includegraphics[width=\linewidth]{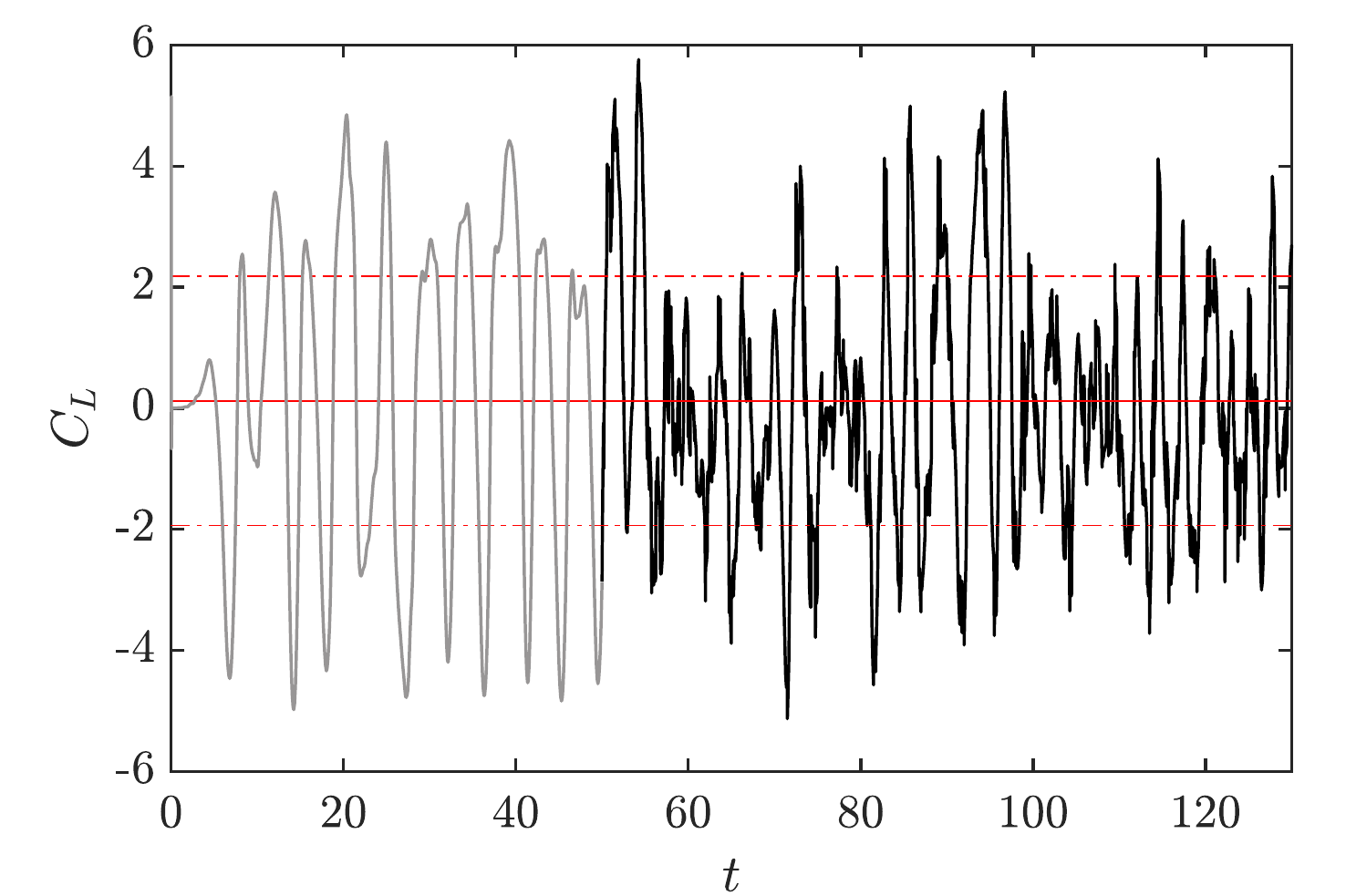}
\captionsetup{justification=centering}
\label{fig:cl_sr_2t2}
\end{subfigure}
\caption{Temporal evolution of $C_D$ (left) and $C_L$ (right) obtained through the application of the DRL control (at $t=50$), run in a deterministic mode, for $Re = 2000$.}
\label{fig:cdcl_sr2k}
\end{figure}

The $C_D$ and $C_L$ obtained through the application of the learning in a deterministic mode is represented in \autoref{fig:cdcl_sr2k}. This figure contrasts sharply with the one observed with $Re = 100$ (\autoref{fig:cdcl_sr}). In this case, the starting baseline flow is much more chaotic, which can be seen in the magnitude of the peaks of both variables. The average drag coefficient of the non-controlled flow is $C_{D}=3.39$. After the control, the average in this period is $C_{D}=2.79$, which means more than \SI{17}{\percent} of $C_D$ improvement. In this case, it has been chosen to represent the mean in the entire controlled period and not during the last vortex-shedding since, as the solution behaves more erratically, it could lead to an average that is far from reality. Regarding the lift coefficient, the applied control manages to set the average $C_{L}$ with a close value to $0$, maximising the reward function as desired and avoiding a systematic biased strategy (see Appendix B in Rabault and Kuhnle \cite{RABAULT2}). Also the absolute value of the peaks during the oscillating periods is reduced but not as much as in the case at Reynolds number $100$, due to the chaotic nature of the flow at this higher Reynolds number.

\begin{figure}[H]
\centering
\subfloat[Average velocity-magnitude fields.]{\includegraphics[height=3.2in]{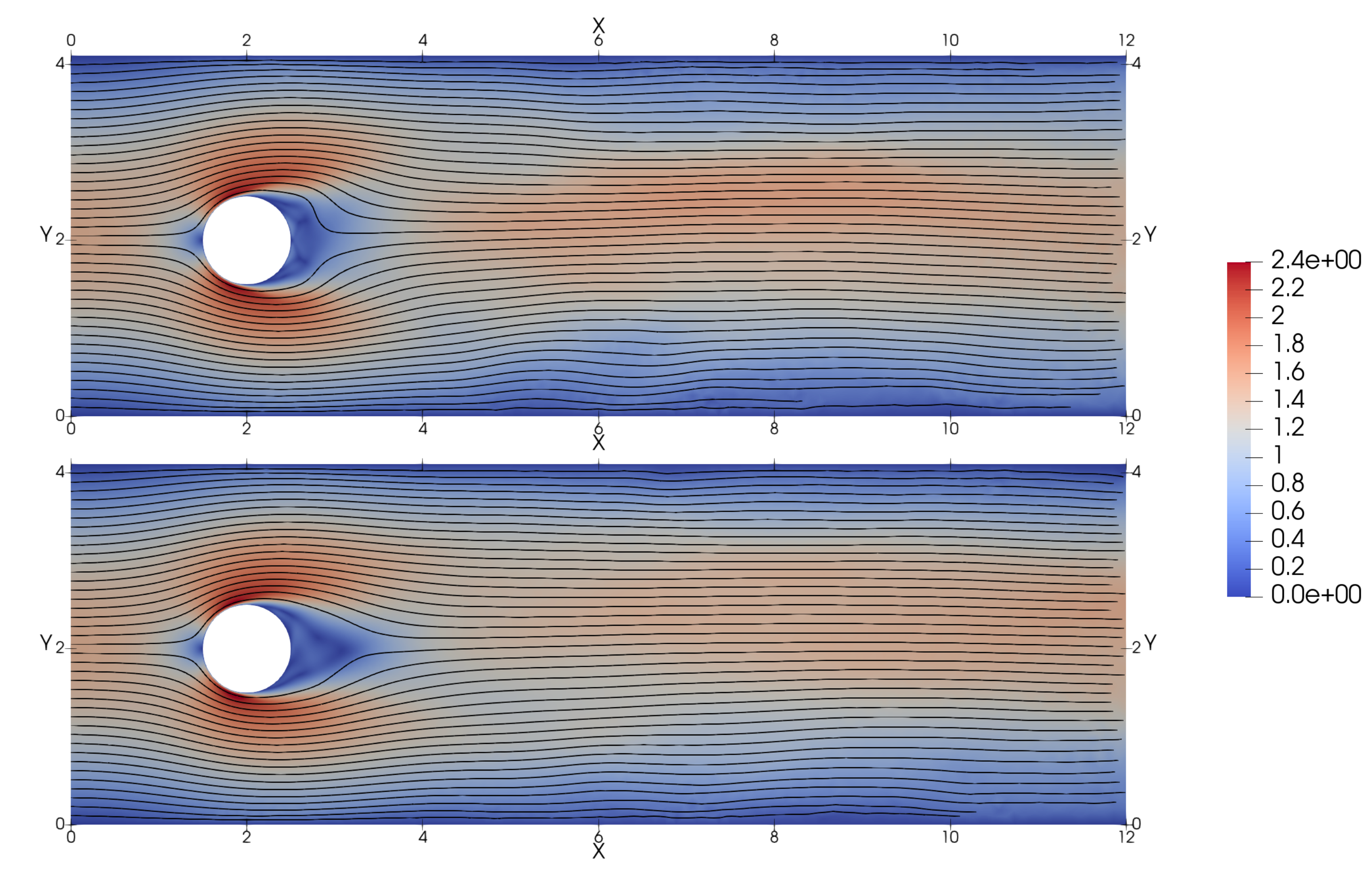}}\\
\vspace{3pt}
\subfloat[Average pressure fields.]{\includegraphics[height=3.2in]{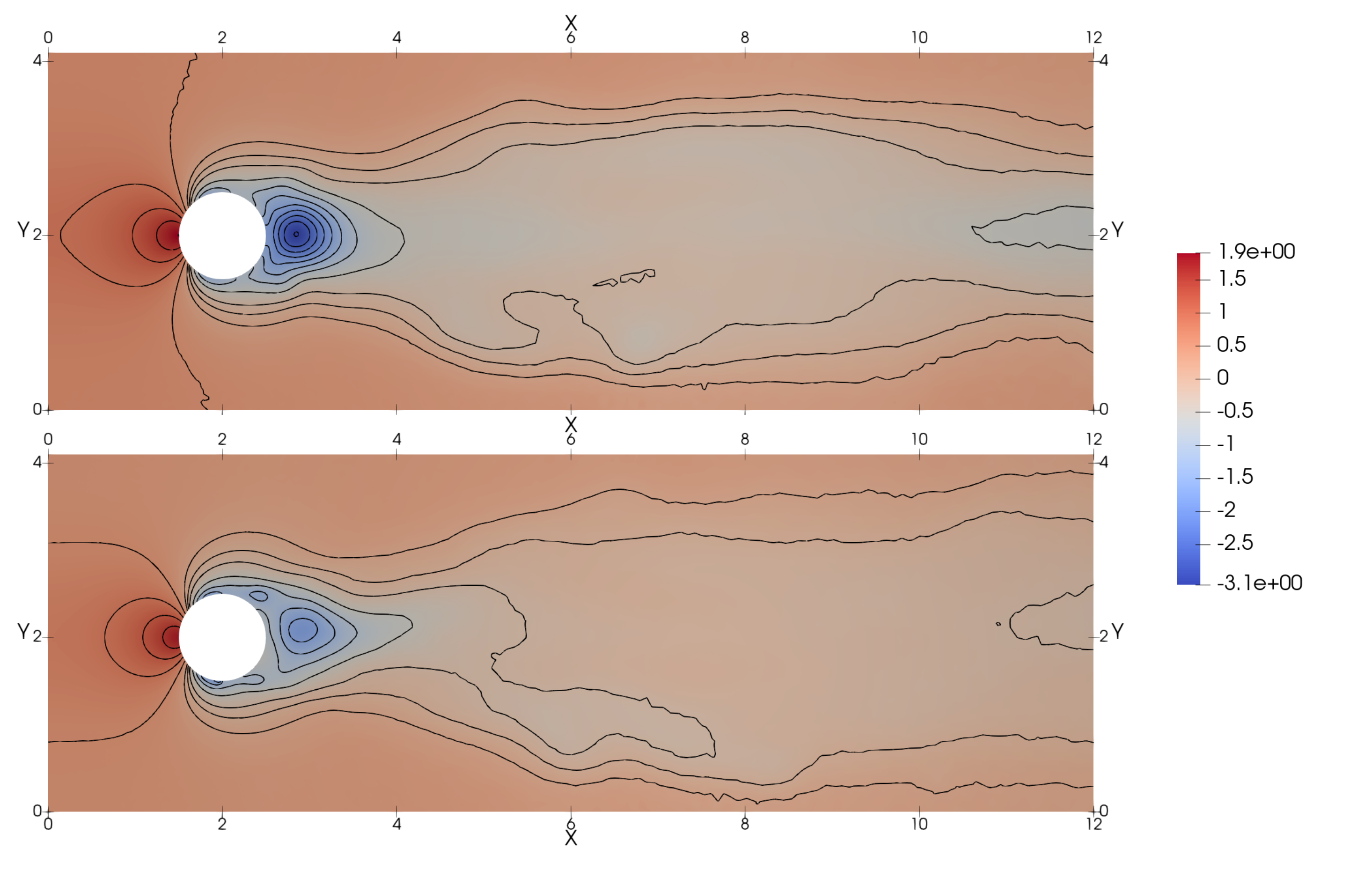}}
\caption{Average flow fields at $Re=2000$, where each pair of images, the baseline case without control is depicted on the top and the controlled case is depicted on the bottom.}
\label{fig:2k_2k}
\end{figure}

\begin{figure}[H]
\centering
\subfloat[Difference in mean velocity-magnitude fields.]{\includegraphics[height=1.6in]{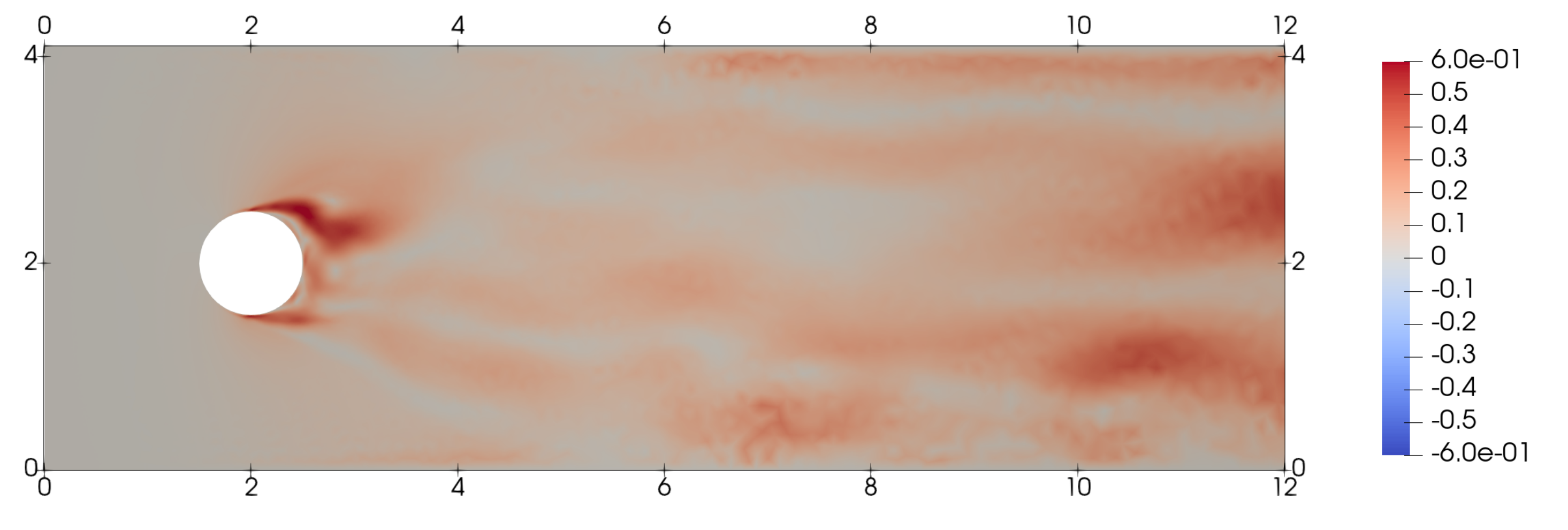}}\\
\vspace{3pt}
\subfloat[Difference in mean pressure fields.]{\includegraphics[height=1.6in]{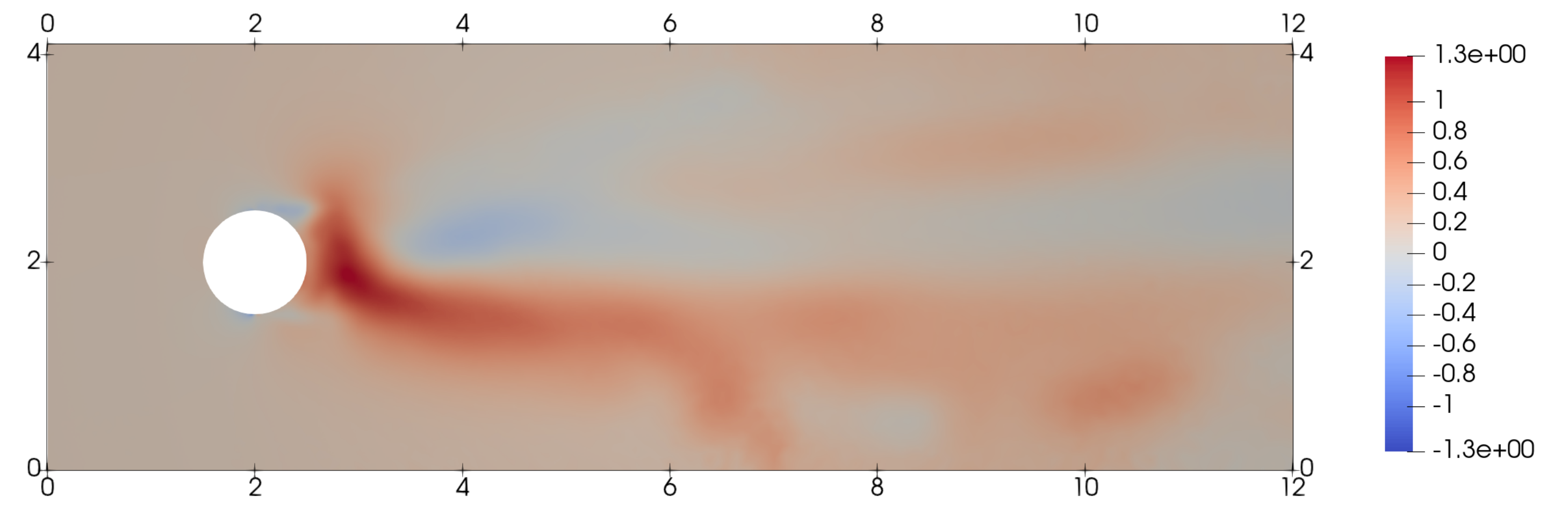}}
\caption{Difference in the mean flow fields at $Re=2000$, between the controlled and baseline case.}
\label{fig:diff2k_2k}
\end{figure}

In \autoref{fig:2k_2k} the average velocity and pressure fields are shown both for the baseline and the controlled deterministic case. As can be seen in the average velocity field comparison, the separation point in the controlled case is further downstream of the cylinder. This way, the average wake is narrowed faster in the controlled case, lowering the drag produced. This phenomenon is easier to observe if the difference between the baseline and the controlled case is considered as shown in \autoref{fig:diff2k_2k}. This figure shows that the more significant difference between the controlled and uncontrolled cases occurs in the separation zone. At the same time, as observed for lower-Reynolds-number control, the pressure drop behind the cylinder is reduced when controlled.

\begin{figure}[H]
\includegraphics[width=1.05\linewidth]{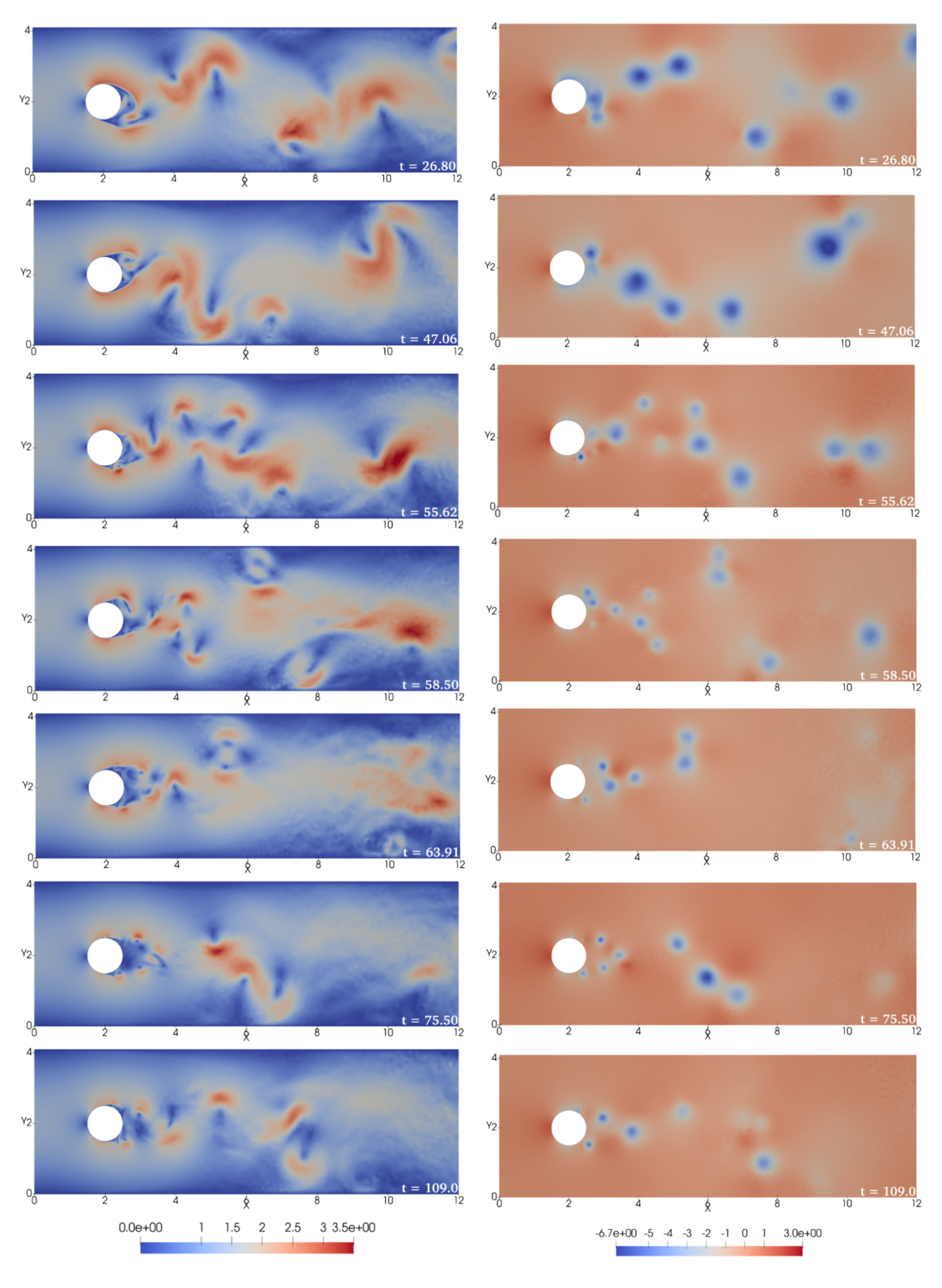}
\caption{Velocity-magnitude (left) and pressure (right) fields progression at $Re=2000$. The control starts at $t=50$, therefore the  first two panels of each column correspond to the uncontrolled case and the rest (involving smaller vortices) correspond to the controlled flow.}
\label{fig:comic2k}
\end{figure}

This control strategy is completely different to that used low at lower Reynolds numbers of values $Re = 100$ and $1000$. In order to better understand this control strategy, a chronological velocity and pressure field snapshot is plotted in \autoref{fig:comic2k}. In this case, the control strategy does not involve the elongation of the recirculation bubble behind the cylinder. The flow separation is controlled and energised by a high-frequency actuation of the jets. The separation point is moved behind the cylinder lowering the drag of the cylinder, similar to the Eiffel paradox phenomenon, also known as the drag crisis, as defined by \citet{DRAGCRISIS}. The agent tries to minimise the drag with this high frequency, breaking the vortices produced after the cylinder into smaller and less-energetic ones.

Additionally, one video of the velocity and pressure fields are shared to aid the visualisation. The corresponding link can be found at the end of the document in the "Data Availability Statement".

\subsection{Cross-application of agents}
\label{cross-t}

Once the $C_D$ improvement results have been obtained for all the calculated Reynolds numbers thanks to the use of DRL, the cross-application of agents for a flow at $Re=2000$ is studied. 

Cross-application of agents involves applying a previously trained ANN with a different Reynolds number flow to solve a similar problem, using the achieved learning in other conditions. The final objective is to reduce the computational cost of training the DRL agent, since training an ANN with a lower Reynolds number is less expensive. This approach can be very beneficial as long as the results produced by the agent are comparable, i.e. the physics are similar enough between the two different cases.

\begin{figure}[H]
\begin{subfigure}[b]{0.48\linewidth}
\includegraphics[width=\linewidth]{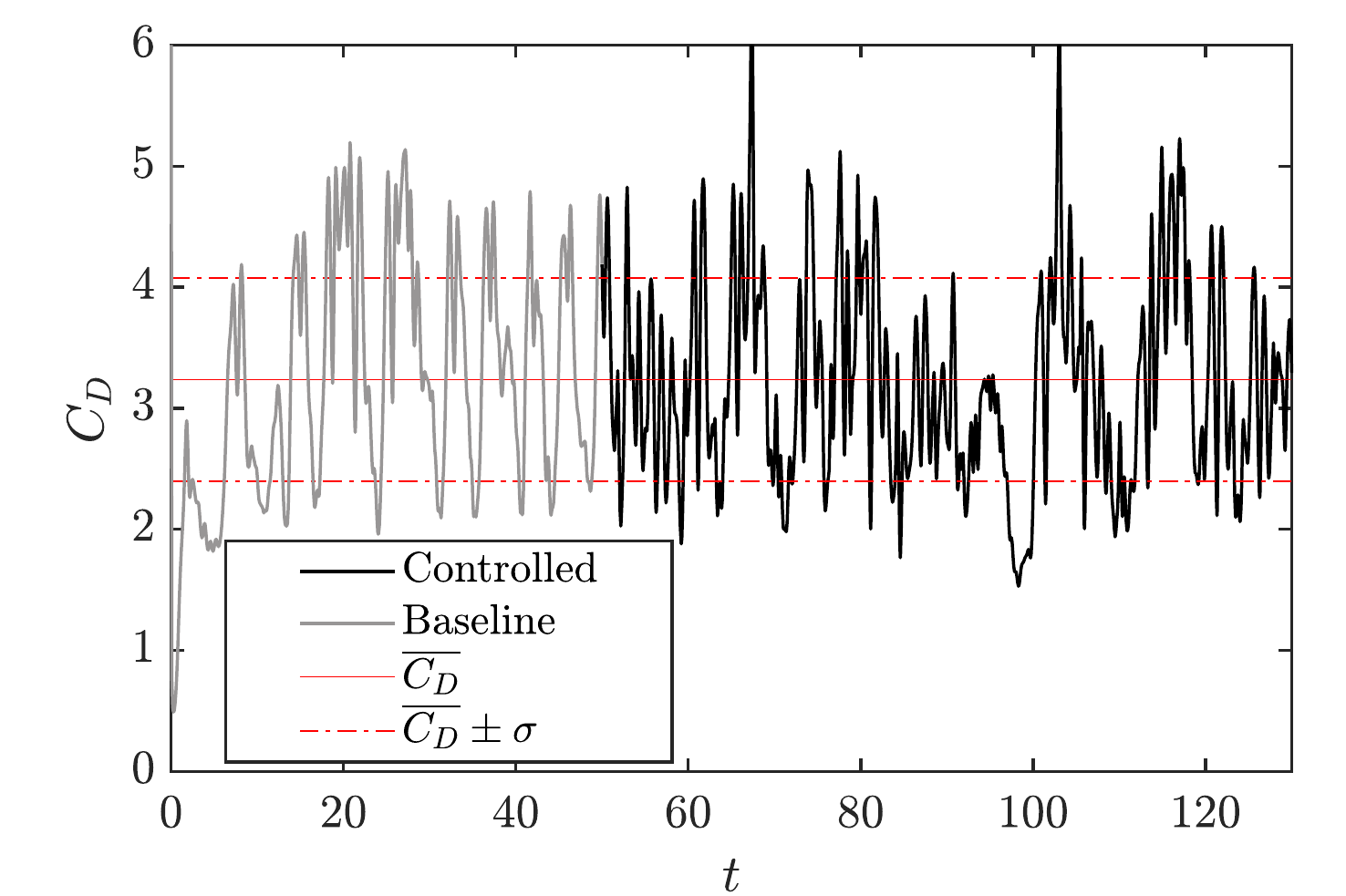}
\captionsetup{justification=centering}
\label{fig:cd_sr2t0}
\end{subfigure}
\begin{subfigure}[b]{0.48\linewidth}
\includegraphics[width=\linewidth]{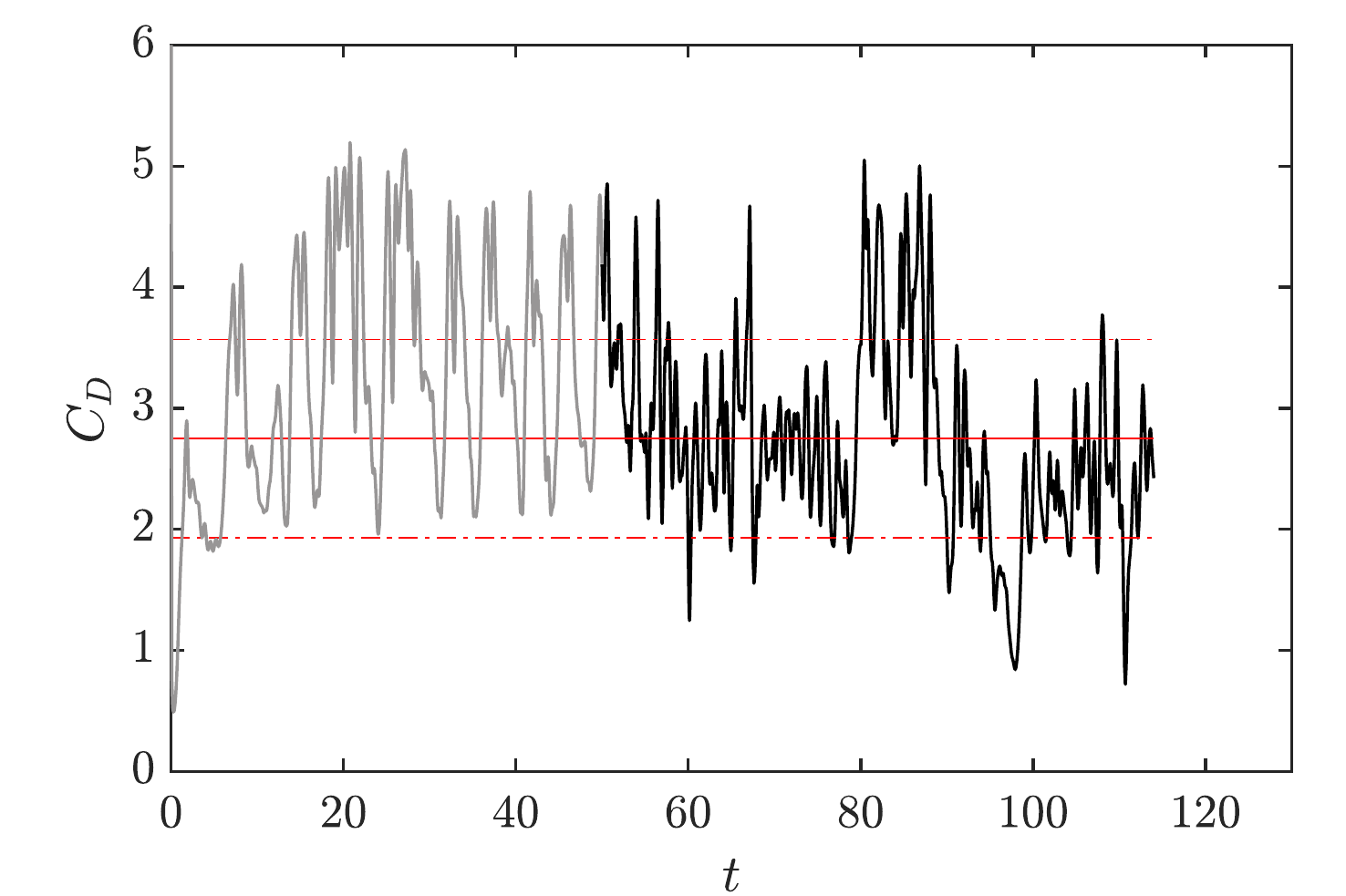}
\captionsetup{justification=centering}
\label{fig:cl_sr_2t1}
\end{subfigure}
\caption{Temporal evolution of $C_D$ obtained through cross-application of agents trained at $Re=100$ (left) and $Re=1000$ (right) in a deterministic mode to a case with a Reynolds number of \num{2000}.}
\label{fig:cd_sr2k_comp}
\end{figure}

Here it is investigated the drag reduction obtained by the agents previously trained for Reynolds numbers \num{100} and \num{1000} when they are applied in a deterministic mode to the flow of $Re=2000$, as can be seen in \autoref{fig:cd_sr2k_comp}.

The cross-application of the agent trained at $Re=100$ does not yield a result that reduces drag in a flow at $Re = 2000$. No drag improvement is obtained because the nature of the flow where the agent is applied is too different to the one where it was trained. However, when the trained agent at $Re = 1000$ is applied for the case of $Re = 2000$, a significant drag reduction is produced. An average $C_{D}$ value of \num{2.74} is obtained, translating into a \SI{19}{\percent} drag reduction improvement compared with the baseline flow. This is a slightly larger drag reduction to that obtained by the agent trained at $Re = 2000$. This small difference can be attributed to the chaotic nature of the flow, from evaluation run to evaluation run, and sub-optimal trade-off with respect to the actual reward function due to less good control of the lift and drag fluctuations, as will be discussed later. The fact that two similar results are obtained with different nature of control strategies, as explained in \autoref{DRL_2000}, may indicate that the flow at $Re = 2000$ belongs to a transition regime towards a higher-Reynolds number flows which only admit a high-frequency control in order to obtain drag reduction. Therefore, being in this transition regime would admit both controls yielding comparable results. In order to go deep into this topic, higher Reynolds numbers must be simulated, but that goes beyond the scope of this article. At the same time, this is a doubly positive result since it confirms the fact that it is possible to apply deep-learning models previously trained at lower Reynolds number but within the same $Re$ regime (similar dynamics) while saving time and computational resources \cite{gua21}. 

\begin{figure}[H]
\begin{subfigure}[b]{0.48\linewidth}
\includegraphics[width=\linewidth]{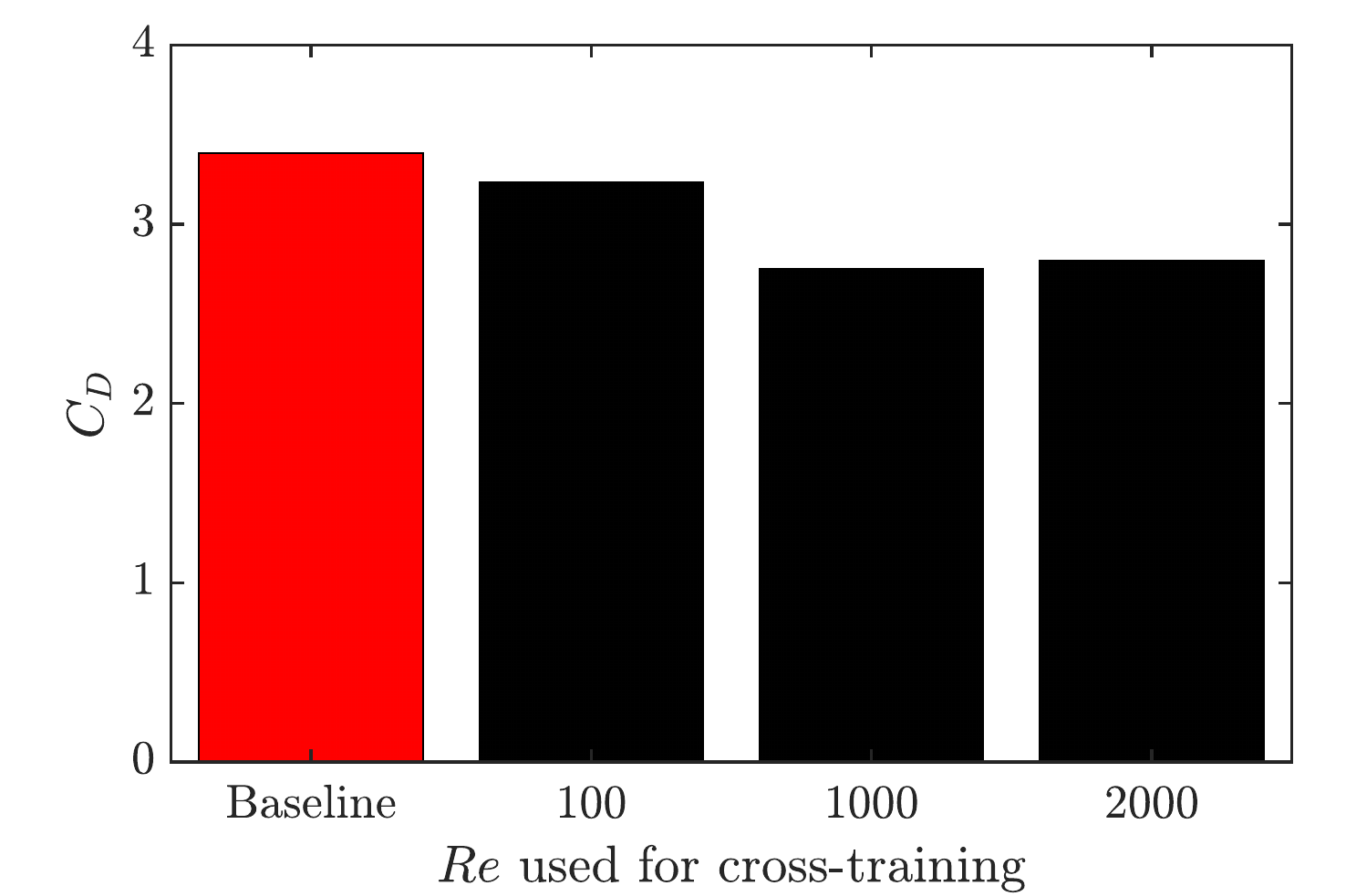}
\captionsetup{justification=centering}
\label{fig:cdcomp1}
\end{subfigure}
\begin{subfigure}[b]{0.48\linewidth}
\includegraphics[width=\linewidth]{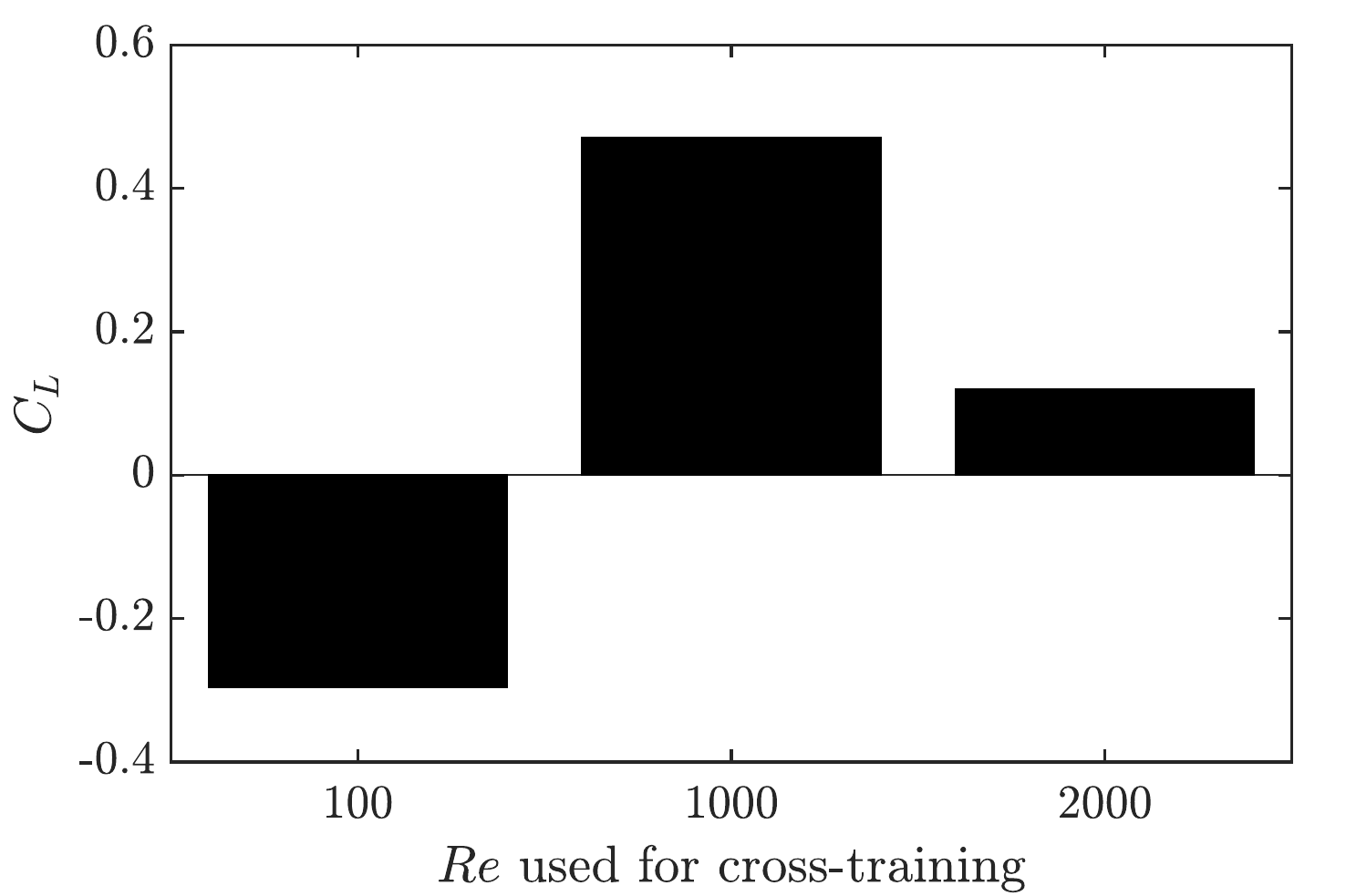}
\captionsetup{justification=centering}
\label{fig:cdcomp2}
\end{subfigure}
\caption{Average $C_D$ (left) and average $C_L$ (right) comparison between three different cross-application of agents.}
\label{fig:cd_imp_comp}
\end{figure}

The average $C_D$ comparisons for each agent used in the cross-application at $Re = 2000$ are shown in \autoref{fig:cd_imp_comp} (left). As detailed in the previous section, the control strategy varies significantly as the Reynolds number increases. While at $Re=100$ and $1000$, the control tries to keep the recirculation bubble behind the cylinder, at $Re=2000$, high-frequency suction and blowing of the jets is applied to quickly force the flow transition, breaking the vortices into smaller ones. Despite these differences in the control strategies between the agents trained at $Re = 1000$ and $2000$, both are capable to obtain a comparable drag reduction in a flow at $Re = 2000$. As shown in \autoref{fig:cd_imp_comp} (right), the $C_L$ bias obtained using the policy from $Re=1000$ is higher than that obtained using the policy of $Re=2000$. Also, $C_L$ fluctuations are larger when employing the agent trained at $Re = 1000$ (not shown here for the sake of brevity). Note that the $Re=2000$ strategy reduces the drag almost as much as using cross-application of agents and, at the same time, exhibits a slightly better performance in the lift; since the lift is part of the reward function, the agent trained at $Re = 2000$ achieves a higher reward at $Re = 2000$ than the agent trained at $Re = 1000$. Nevertheless, the small differences between both trainings may be reversed for different baseline flows (changing the initial condition). This is similar to the results observed by \citet{REN}.

 %This may indicate that the typical range of Reynolds number between \num{1000} and \num{2000} is where the transition between an optimal strategy corresponding to opposition wake control and an strategy consisting in the separation of the area of energisation is taking place.

\begin{figure}[H]
\centering
\includegraphics[height=1.33in]{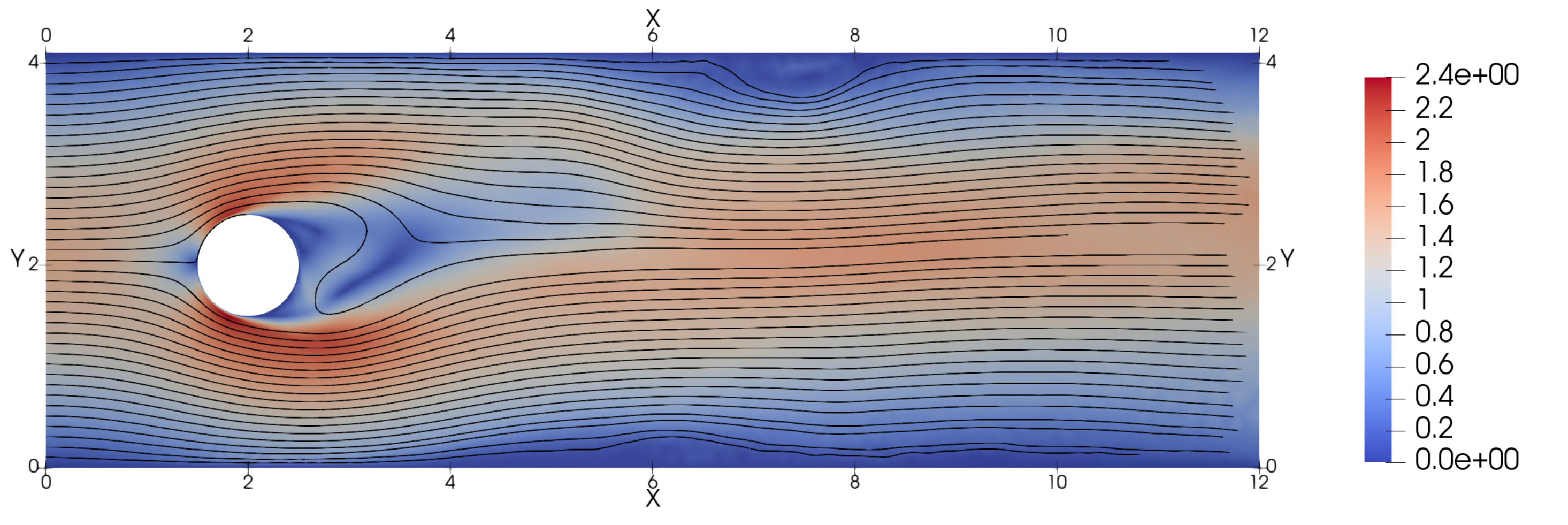}
\caption{Average flow fields at $Re=2000$, using the training obtained for $Re=1000$.}
\label{fig:2k_1k}
\end{figure}

In \autoref{fig:2k_1k}, the average velocity-magnitude field at $Re=2000$ is shown for the cross-application of the agent trained at $Re=1000$. At first glance, it is noticeable the existent bias in the average velocity field, which is directly linked with the asymmetric lift behaviour.

\begin{figure}[H]
\begin{subfigure}[b]{0.48\linewidth}
\includegraphics[width=\linewidth]{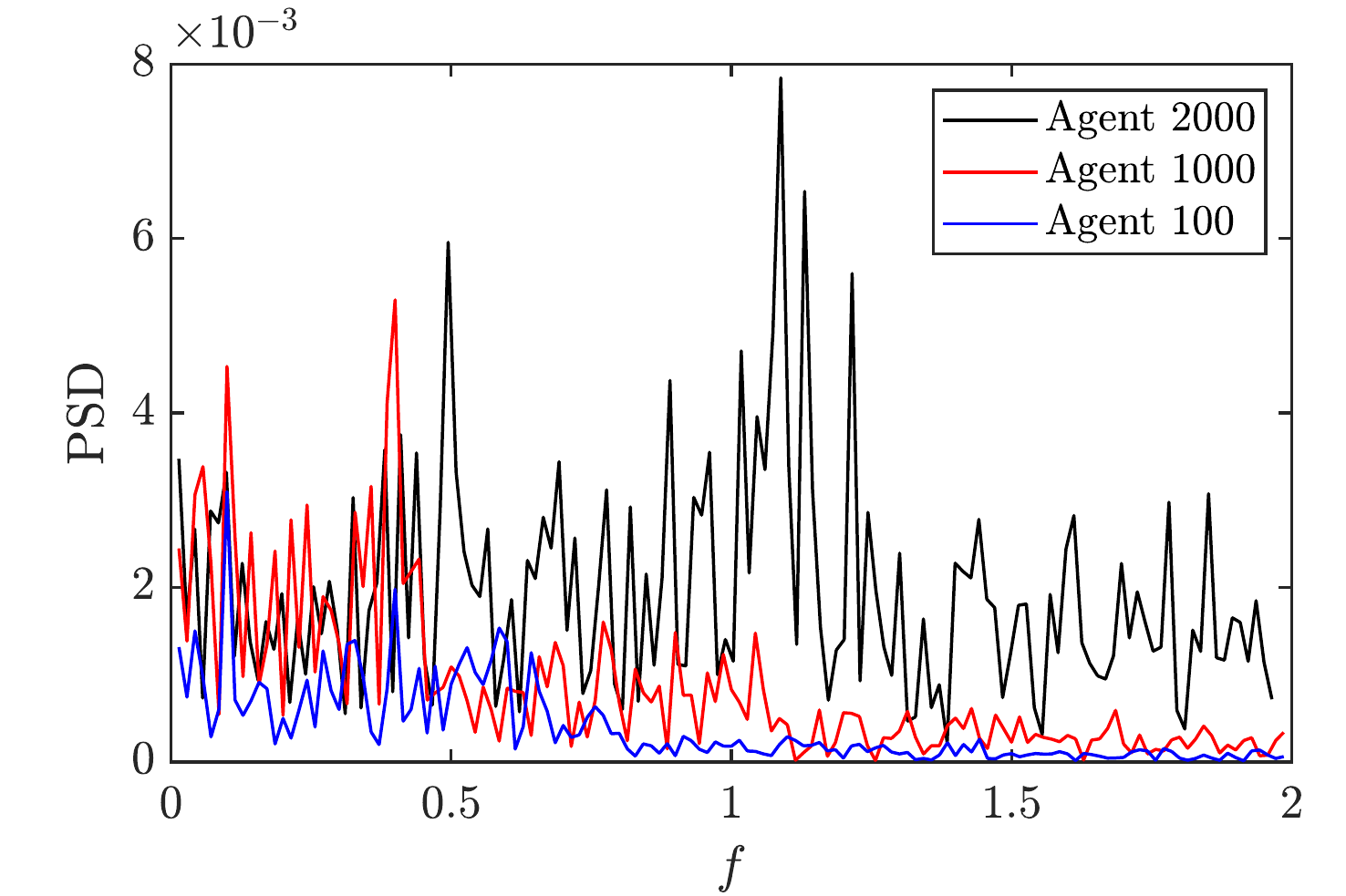}
\captionsetup{justification=centering}
\label{fig:pn_f}
\end{subfigure}
\begin{subfigure}[b]{0.48\linewidth}
\includegraphics[width=\linewidth]{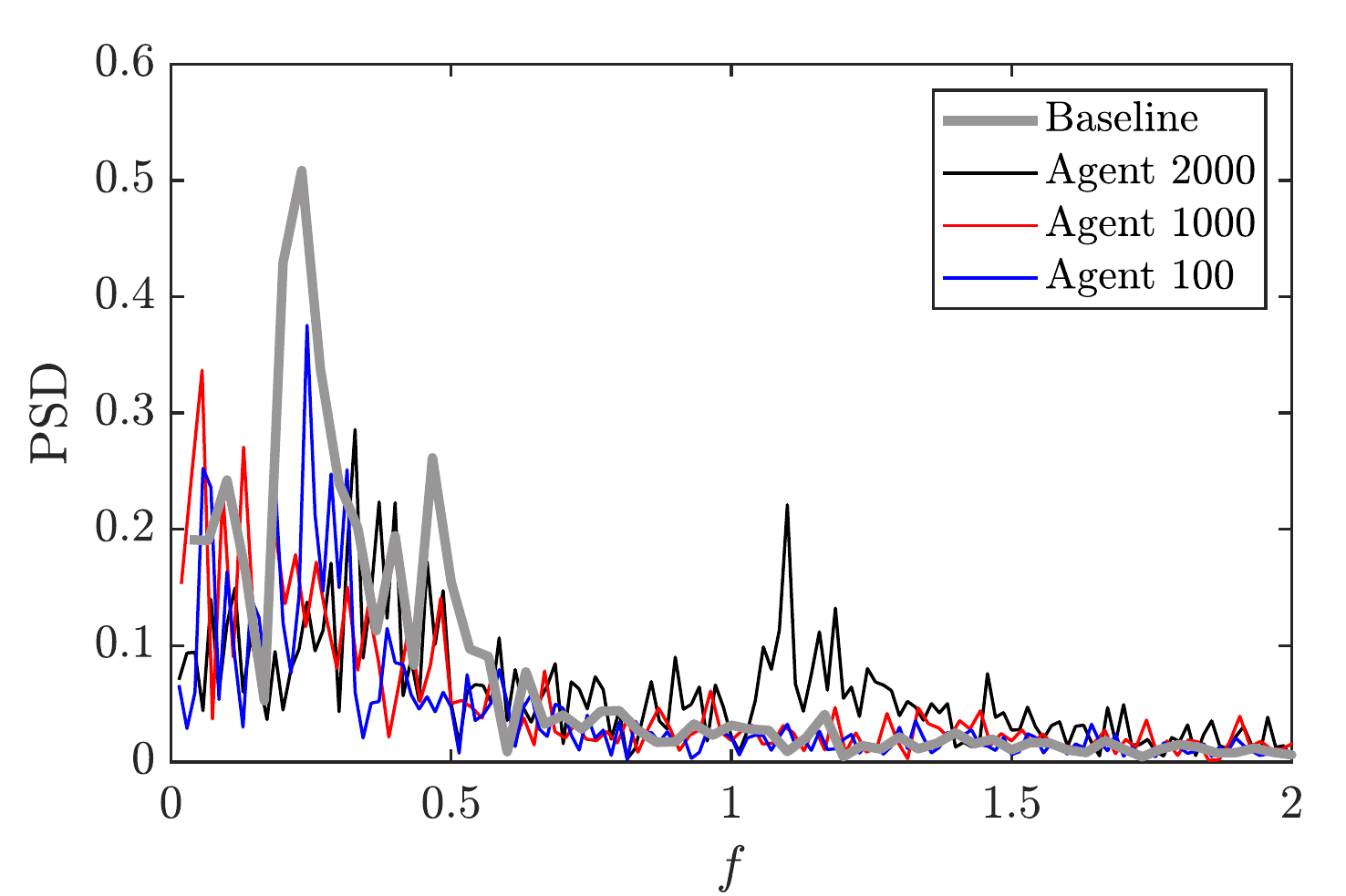}
\captionsetup{justification=centering}
\label{fig:pf_f}
\end{subfigure}
\caption{Power-spectral density the action output, $Q$, for the $Re = 2000$ case, with its own policy and two cross-application policies (left) and for pressure in the probe near the cylinder in the separation zone (right).}
\label{fig:p_f}
\end{figure}

In order to quantitatively observe the different control strategies, the power-spectral density (PSD) of the actions is plotted for the policy of $Re=2000$, and the cross-application of agents, $Re=1000$ and $100$, in \autoref{fig:p_f} (left). The latter two agents share the same strategy, which consist of a low-frequency action with no significant content at medium and high frequencies. In contrast, with the policy of $Re=2000$, the frequency spectrum is distributed, with a peak at a frequency of $1.1$. In \autoref{fig:p_f} (right), the frequency spectrum of the pressure at the probe in the detachment region indicated in \autoref{fig:witness} is shown. A low-frequency of $0.24$ governs the baseline flow, corresponding to the vortex-shedding frequency. This frequency is mitigated by the policy of $Re=2000$ in higher frequencies. Specifically, the frequency peak at $1.1$ is a consequence of the corresponding actuations at this frequency, which is the responsible of breaking the flow into smaller vortices as seen in \autoref{fig:comic2k}.
On the other hand, it can be seen how the agent trained at $Re = 1000$ tends towards a completely different strategy in terms of frequency, actuating with a low-frequency strategy. The agent trained at $Re = 100$, also uses this low-frequency actuation, but it does not have the capabilities to influence the pressure field. As mentioned before, in those cases, the agent tries an opposition control strategy increasing the recirculation bubble to minimise the drag.

\section{Conclusions}\label{sec:Conclusions}

In this work, the high-performance CFD solver Alya has been coupled with an ANN DRL agent to simulate and control the flow around a 2D cylinder with two jets attached to the cylinder surface. The main control objective has been set on the reduction of the average drag. For the first time, the Reynolds number of the canonical 2D cylinder control problem from \citet{RABAULT1} has been extended to $Re=2000$ , providing new insights into the DRL control strategies for highly complex and dynamical flows. Additionally, the $Re=100$ and $Re=1000$ cases have been studied for validation and comparison purposes.

The most striking outcome of  case is that the DRL agent uses a radically different strategy from those obtained at lower Reynolds numbers while being able to provide a \SI{17}{\percent} drag reduction. In the new strategy, the agent tries to delay the detachment point in the cylinder surface using a high-frequency signal in the actuation of the jets, similarly to what can be observed in drag crisis phenomena. It is shown that the cylinder wake is narrowed by the breakdown of the detaching vortices into smaller structures. This strategy contrasts with the one obtained at a lower Reynolds number, where the agent acts at a lower frequency to perform opposition control and to elongate the recirculation bubble behind the cylinder. These results are further verified with the spectral analysis of the jet actuating signals and a pressure witness point in the wake.

Finally, the application of agents trained at $Re=100$ and $Re=1000$ on the $Re=2000$ case has been investigated (namely cross-application). It has been shown that the $Re=100$ agent is not able to reduce the drag in the high Reynolds regime, due to the different dynamics of the system. On the other hand, the $Re=1000$ provides satisfactory results which even beat the $Re=2000$ agent itself (\SI{19}{\percent} drag reduction), even though the mean wake displays an asymmetric pattern in this case, which corresponds to an overall reward (including lift bias penalisation) that is lower than with the agent trained at $Re=2000$. Still, this opens the door to accelerate the training of an agent by first exposing it to a lower Reynolds-number flow, which demands a lower computational effort, and then finalise the training at the target Reynolds-number condition.

% separate names by semicolons   
\authorcontributions{%
Conceptualisation: RV, JR, OL; Investigation: all authors; Data curation: PV, PS, FAA, AM, JR, BF; Software: JR, AM, BF, OL; Writing original draft: PV, PS, FAA; Editing: all authors; Funding: LMGC, OL, RV. All authors have read and agreed to the published version of the manuscript.
}

\funding{Pau Varela is partially supported through a grant for the mobility of doctoral students provided by Universitat Politècnica de València and the program Erasmus Prácticas E+ 2020-1. RV acknowledges funding by the ERC through Grant No. ``2021-CoG-101043998, DEEPCONTROL''.}

\institutionalreview{Not applicable.}

\informedconsent{Not applicable.}

\dataavailability{The data presented in this study are available on request from
the corresponding author.

YouTube link to "Deep reinforcement learning for flow control on a cylinder a Re=2,000":

\url{https://youtu.be/8R3adCQmeEA}}

\acknowledgments{The authors acknowledge the contribution of Maxence Deferrez to this work. RV acknowledges funding by the ERC through Grant No. ``2021-CoG-101043998, DEEPCONTROL''.}

\conflictsofinterest{The authors declare no conflict of interest. The funders had no role in the
design of the study; in the collection, analyses, or interpretation of data; in the writing of the
manuscript, or in the decision to publish the results.} 

%%%%%%%%%%%%%%%%%%%%%%%%%%%%%%%%%%%%%%%%%%

\abbreviations{The following abbreviations are used in this manuscript:\\

  \label{Nom}
  \begin{supertabular}{@{}ll}
   \multicolumn{2}{l}{Abbreviations} \\
    AFC & Active flow control \\
    ANN & Artificial neural network \\
    BSC-CNS & Barcelona Supercomputing Center - Centro Nacional de Supercomputación \\
    CFD & Computational fluid dynamics \\
    CFL & Courant–Friedrichs–Lewy \\
    CPU & Central processing unit \\
    DRL & Deep reinforcement learning \\
    EMAC & energy-, momentum- and angular-momentum-conserving equation\\
    FEM & Finite-element method \\
    HPC & High-performance computing \\
    PPO & Proximal-policy optimisation \\
    PSD & Power-spectral density \\
    UAV & Unmanned aerial vehicle\\

   & \\
   \multicolumn{2}{l}{Roman letters} \\
     $a$ & action \\
   $C_{L}$ & Lift coefficient \\
	 $C_{D}$ & Drag coefficient \\
   $C_{\rm{offset}}$ & Offset coefficient of the reward \\
   $D$ & Cylinder diameter \\
   $e_{j}$ & Vector used in force calculation\\
   $f$ & External forces, frequency\\
   $f_{k}$ & Vortex shedding frequency\\
   $F$ & Force\\
   $F_{D}$ & Drag force\\
   $F_{L}$ & Lift Force\\
   $H$ & Channel height\\
   $L$ & Channel length\\
   $n$ & Unit vector normal to the cylinder\\
   $Q$ & Mass flow rate\\
    $Q^{*}$ & Normalised mass flow rate\\
    $Q_\textnormal{ref}$ & Reference mass flow rate\\
    $p$ & Pressure\\
   $r$ & Reward \\
   $r_{\rm{norm}}$ & Reference value of the reward after control \\
    $R$ & Cylinder radius \\
	 $Re$ & Reynolds number \\
	 $S$ & Surface \\
	 $s$ & observation state \\
	 $s_{\rm{norm}}$ & Reference pressure in the observation state \\
	 $St$ & Strouhal number \\
	 $t$ & Time \\
	 $t_{0}$ & Initial time \\
	 $t_f$ & Final time \\
	 $T_{a}$ & Action period \\
	 $T_{k}$ & Vortex-shedding period \\
	 $\vb{u}$ & Flow speed \\
	 $\bar{U}$ & Mean velocity \\
   $U_\textnormal{in}$ & Inlet boundary velocity in x direction\\
   $U_\textnormal{max}$ & Inlet boundary velocity in the middle of the channel \\
   $V_\textnormal{in}$ & Inlet boundary velocity in y direction\\
   $v_i$ & Jet velocity \\
      $w$ & Lift penalisation \\
   $x$ & Horizontal coordinate \\
   $y$ & Vertical coordinate\\

   & \\
      & \\
   \multicolumn{2}{l}{Greek letters} \\
     $\epsilon$ & Velocity strain-rate tensor \\
     $\nu$ & Kinematic viscosity \\
     $\Omega$ & Domain \\
     $\omega$ & Jet angular opening \\
	 $\rho$ & Density \\
	 $\sigma$ & Standard deviation \\
	 $\varsigma$ & Cauchy stress tensor\\
	 $\theta$ & Jet angle\\
	 $\theta_{0}$ & Centre jet angle\\
  \end{supertabular}}

\end{paracol}
\reftitle{References}

% Please provide either the correct journal abbreviation (e.g. according to the ``List of Title Word Abbreviations'' http://www.issn.org/services/online-services/access-to-the-ltwa/) or the full name of the journal.

\externalbibliography{yes}
\bibliography{bibliography.bib}

%=====================================
% References, variant B: internal bibliography
%=====================================

% If authors have biography, please use the format below
%\section*{Short Biography of Authors}
%\bio
%{\raisebox{-0.35cm}{\includegraphics[width=3.5cm,height=5.3cm,clip,keepaspectratio]{Definitions/author1.pdf}}}
%{\textbf{Firstname Lastname} Biography of first author}
%
%\bio
%{\raisebox{-0.35cm}{\includegraphics[width=3.5cm,height=5.3cm,clip,keepaspectratio]{Definitions/author2.pdf}}}
%{\textbf{Firstname Lastname} Biography of second author}

% The following MDPI journals use author-date citation: Arts, Econometrics, Economies, Genealogy, Humanities, IJFS, JRFM, Laws, Religions, Risks, Social Sciences. For those journals, please follow the formatting guidelines on http://www.mdpi.com/authors/references
% To cite two works by the same author: \citeauthor{ref-journal-1a} (\citeyear{ref-journal-1a}, \citeyear{ref-journal-1b}). This produces: Whittaker (1967, 1975)
% To cite two works by the same author with specific pages: \citeauthor{ref-journal-3a} (\citeyear{ref-journal-3a}, p. 328; \citeyear{ref-journal-3b}, p.475). This produces: Wong (1999, p. 328; 2000, p. 475)

%%%%%%%%%%%%%%%%%%%%%%%%%%%%%%%%%%%%%%%%%%
%% for journal Sci
%\reviewreports{\\
%Reviewer 1 comments and authors' response\\
%Reviewer 2 comments and authors' response\\
%Reviewer 3 comments and authors' response
%}
%%%%%%%%%%%%%%%%%%%%%%%%%%%%%%%%%%%%%%%%%%

\end{document}